%% file: main.tex
\documentclass[preprint,12pt]{elsarticle}

\usepackage{amssymb}
\usepackage{graphicx}%
\usepackage{multirow}%
\usepackage{amsmath, amsfonts}%
\usepackage{amsthm}%
\usepackage{mathrsfs}%
\usepackage{xcolor}%
\usepackage{textcomp}%
\usepackage{manyfoot}%
\usepackage{booktabs}%
\usepackage{algorithm}%
\usepackage{algorithmicx}%
\usepackage{algpseudocode}%
\usepackage{listings}%
\usepackage{nameref}
\usepackage{lscape}
\usepackage{adjustbox}
\usepackage[hidelinks]{hyperref}
\hypersetup{
    colorlinks,
    linkcolor={red!50!black},
    citecolor={blue!50!black},
    urlcolor={blue!80!black}
}
\usepackage[shortlabels]{enumitem}
\usepackage{threeparttable}
\usepackage{subcaption}
\usepackage{tabularx}
\usepackage{rotating}
\usepackage{array}

\usepackage{kotex}

\colorlet{r1}{red}
\colorlet{r2}{teal}
\colorlet{r3}{orange}
\colorlet{r5}{magenta}
\colorlet{a}{blue}

\journal{Biomedical Signal Processing and Control}

\begin{document}

\begin{frontmatter}



\title{
Personalized Deep Learning for Short-Term Forecasting of Impending Atrial Fibrillation from Continuous Wearable ECG Signals
}


\author[inst1]{Jangwon Suh\corref{eq}}
\author[inst2]{Soonil Kwon\corref{eq}}
\author[inst3]{Jungmin Ko}
\author[inst4]{Yun Kwan Kim}
\author[inst4]{Hee Seok Song}
\author[inst5]{Eue-Keun Choi\corref{cor1}}
\author[inst1,inst3]{Wonjong Rhee\corref{cor1}}

\affiliation[inst1]{organization={Department of Intelligence and Information, Seoul National University},
            addressline={1, Gwanak-ro, Gwanak-gu},
            city={Seoul},
            postcode={08826},
            country={Republic of Korea}}

\affiliation[inst2]{organization={Division of Cardiology, Department of Internal Medicine, CHA Bundang Medical Center and CHA University School of Medicine},
            addressline={59, Yatap-ro, Bundang-gu},
            city={Seongnam},
            postcode={13496},
            country={Republic of Korea}}

\affiliation[inst3]{organization={Interdisciplinary Program in Artificial Intelligence, Seoul National University},
            addressline={1, Gwanak-ro, Gwanak-gu}, 
            city={Seoul},
            postcode={08826}, 
            country={Republic of Korea}}

\affiliation[inst4]{organization={SEERS Technology Co., Ltd},
            addressline={291-13, Dongbu-daero, Jinwi-myeon},
            city={Pyeongtaek},
            postcode={17707},
            country={Republic of Korea}}

\affiliation[inst5]{organization={Department of Internal Medicine, Seoul National University College of Medicine and Seoul National University Hospital},
            addressline={101, Daehak-ro, Jongno-gu},
            city={Seoul},
            postcode={03080},
            country={Republic of Korea}}

\cortext[eq]{These authors contributed equally to this work}
\cortext[cor1]{Corresponding authors: E.-K. Choi, choiek417@gmail.com \& W. Rhee, wrhee@snu.ac.kr}

\begin{abstract}
\input{TEX/00.abstract}
\end{abstract}

% \includegraphics[width=\textwidth]{figures/framework_v6.pdf}

% \begin{highlights}
% \item A large-scale assessment of feature attribution methods is provided.
% \item Eleven feature attribution methods are considered over five large ECG datasets.
% \item Both automatic and human evaluations are performed.
% \item In our experiments, Guided Grad-CAM exhibited outstanding performance.
% \end{highlights}


\end{frontmatter}


\input{TEX/10.introduction}
\input{TEX/20.related_works}
\input{TEX/30.methods}
\input{TEX/40.results}
\input{TEX/50.discussion}
\input{TEX/60.conclusion}
\input{TEX/70.additional}

\clearpage
\appendix

\input{TEX/A1.implementation_details}\clearpage
\input{TEX/A2.implementation_of_feature_attribution_analysis}\clearpage
\input{TEX/A3.AF_annotation_lengths}\clearpage
\input{TEX/A4.comparison_with_personalization_from_scratch}\clearpage
\input{TEX/A5.additional_tables}

\clearpage

\bibliographystyle{elsarticle-num-names}
\bibliography{cas-refs}





% \end{thebibliography}
\end{document}

%% file: TEX/00.abstract.tex
\textbf{Background and Objective:}
% Continuous wearable electrocardiogram (ECG) monitoring is increasingly used for ambulatory arrhythmia surveillance, yet forecasting impending atrial fibrillation (AF) from these signals remains a significant challenge, as the performance of signal-analysis models is often hindered by inter-patient variability in ECG morphology. This study investigated whether personalizing a global model---trained on multi-patient recordings---by fine-tuning it on an individual's own ECG signals could improve short-term forecasting of impending AF for real-time patient monitoring.
Continuous wearable electrocardiogram (ECG) monitoring is increasingly used for ambulatory arrhythmia surveillance, yet forecasting impending atrial fibrillation (AF) is challenged by inter-patient ECG variability. This study investigated whether personalizing a global model via fine-tuning on an individual's ECG signals improves short-term forecasting of impending AF.

\noindent \textbf{Methods:}
% A global forecasting model was initially trained on the ICENTIA11K dataset.
% This model was then compared against personalized models fine-tuned across three distinct cohorts: ICENTIA11K, IRIDIA-AF, and MobiCARE.
% After a signal-quality and preprocessing pipeline (baseline-wander removal, normalization, and noise rejection), all models processed 60-second ECG segments to target a five-minute forecast horizon for impending AF.
% The study also evaluated the impact of the number of patient-specific episodes used for adaptation and analyzed ECG features, such as heart rate and root mean square of successive differences (RMSSD).
A global model trained on the ICENTIA11K dataset was compared against personalized models fine-tuned across three cohorts: ICENTIA11K, IRIDIA-AF, and MobiCARE. Following preprocessing, models processed 60-second ECG segments for a five-minute forecast horizon. We evaluated the impact of adaptation data volume and analyzed ECG features, such as heart rate and RMSSD.

\noindent \textbf{Results:}
% Personalized models demonstrated significant performance gains over the global model in two cohorts, achieving an AUROC of 0.711 vs. 0.614 in ICENTIA11K and 0.686 vs. 0.585 in MobiCARE.
% The benefits of personalization generally increased with the number of patient-specific episodes available for fine-tuning.
% While the global model's accuracy rose as AF onset approached, personalized models in the two external cohorts exhibited distinct temporal dynamics, which may indicate the capture of patient-specific characteristics less dependent on proximity to the AF event.
% Analysis of ECG features confirmed that pre-AF episodes showed significantly higher heart rates and RMSSD.
% Feature attribution analysis indicated that the model focused on clinically relevant arrhythmic precursors, including frequent premature atrial complexes (PACs) and short supraventricular tachycardias (SVTs).
Personalized models significantly outperformed the global model, achieving AUROCs of 0.711 vs. 0.614 in ICENTIA11K and 0.686 vs. 0.585 in MobiCARE. Personalization benefits increased with the amount of patient-specific fine-tuning data. While the global model's accuracy rose as AF onset approached, personalized models in the two external cohorts exhibited distinct temporal dynamics, which may indicate the capture of patient-specific characteristics less dependent on proximity to the AF event. Pre-AF episodes showed elevated heart rates and RMSSD. Feature attributions highlighted clinically relevant precursors, including frequent premature atrial complexes (PACs) and short supraventricular tachycardias (SVTs).

\noindent \textbf{Conclusions:}
% Adapting deep learning models with patient-specific wearable ECG data significantly enhances the short-term forecasting of impending AF. This personalized framework supports timely preventive interventions and improved management of AF within ambulatory monitoring environments.
Adapting deep learning models with patient-specific wearable ECG data significantly enhances short-term forecasting of impending AF. This personalized framework supports timely preventive interventions and improved AF management in ambulatory monitoring environments.

%% file: TEX/10.introduction.tex
\section{Introduction}\label{sec:introduction}

Atrial fibrillation (AF) is the most common cardiac arrhythmia, not only increasing ischemic stroke, heart failure, and mortality, but also continuing to increase its prevalence globally~\citep{linz2024atrial, van20242024}. Because early diagnosis and management are crucial for preventing disease progression and improving prognosis~\citep{kirchhof2020early, schnabel2023early}, modern ambulatory electrocardiogram (ECG) monitoring tools such as wearable ECG patches have been introduced into clinical practice~\citep{svennberg2022use}. Technological advances have led to compact and wireless ambulatory ECG platforms, expanding their use in long-term rhythm surveillance and improving the detection of paroxysmal or asymptomatic AF episodes~\citep{steinhubl2018effect, ha2021effect, kwon2021validation, kwon2022comparison}.

Although the forecasting of paroxysmal AF using retrospective analysis of non-AF ECG has been well demonstrated~\citep{attia2019artificial, lee2024mobile}, a clinically unmet need is to forecast the AF onset or impending AF episodes. Short-term forecasting of AF may help highly symptomatic patients prevent AF attacks by providing early warning and enabling a `pill-in-the-pocket' strategy before AF onset. Recent studies have investigated deep learning approaches to predict AF onset from minutes to weeks. \citet{gavidia2024early} predicted AF onset approximately 30 minutes in advance. \citet{singh2022short} utilized 24-hour Holter recordings to predict AF risk over the subsequent two weeks. Similarly, \citet{gadaleta2023prediction} demonstrated that single-lead ECG monitoring could predict impending AF within two weeks. However, performance in these population-level models may be constrained by inter-individual variability in ECG characteristics, including differences in P-wave, QRS complex, ST segment, and T-wave morphologies. Accordingly, we hypothesize that global models---trained on pooled data across many patients---may have limited predictive power, whereas personalized models could provide superior performance.

\begin{figure}[htbp]
\centering
\includegraphics[width=\textwidth]{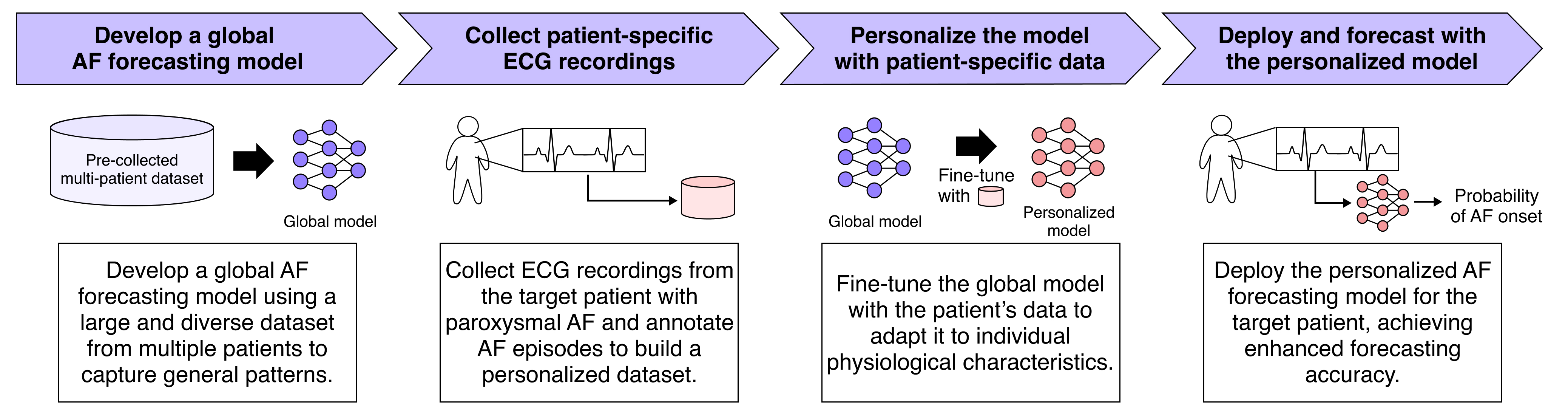}
\caption{Workflow of personalized AF forecasting. With the advent of ambulatory ECG monitoring devices, real-time ECG data have become readily available, enabling the development of personalized short-term AF forecasting models for these systems.}
\label{fig:figure1-1}
\end{figure}

Unlike brief, intermittent ECG recordings, continuous ECG data collected by ambulatory monitors provide abundant, patient-specific context about an individual’s cardiac rhythms.
In particular, segments from an individual's own recordings can fine-tune a global model into a patient-adapted model, tailoring it to each patient's unique ECG patterns to improve forecasting accuracy.
Figure~\ref{fig:figure1-1} illustrates the workflow for personalized AF forecasting.

In this study, we investigate a personalized signal-analysis framework for forecasting impending AF episodes from continuous single-lead ECG acquired by a wearable monitoring device, aimed at supporting real-time alerting and patient management in ambulatory settings, with potential extension to critical-care monitoring.
Specifically, our main contributions are:

\begin{itemize}

\item We formalize and compare \textbf{global (population-level)} versus \textbf{personalized (patient-level)} models for short-term AF forecasting from single-lead wearable ECG signals, targeting a five-minute horizon suited to real-time monitoring and timely intervention.

\item We run a multi-cohort evaluation on three single-lead ECG datasets (ICENTIA11K, IRIDIA-AF, and MobiCARE). Personalized fine-tuning generally improves performance over the global model, with statistically significant gains in two cohorts and numerical improvements in the third.

\item We study data needs for personalization by varying the number of patient episodes used for adaptation, finding larger gains with more episodes and a risk of overfitting when only a few are available. This motivates a data-aware policy for enabling personalization.

\item We analyze forecasting performance by temporal distance to AF onset. The global model’s accuracy rises near onset, while personalized models show different temporal dynamics, which may reflect patient-specific cues that enable earlier forecasting of impending AF.

\item We provide interpretability via physiological summaries and feature attributions. Episodes preceding AF exhibit elevated heart rate and RMSSD, and the model’s feature attributions emphasize clinically plausible precursors such as a greater burden of premature atrial complexes and short supraventricular tachycardia episodes.
\end{itemize}

Overall, this study demonstrates that personalized deep learning models based on ambulatory ECG monitoring can improve short-term AF forecasting. Such a framework may help physicians and patients better manage AF through a personalized approach in digital medicine.

%% file: TEX/20.related_works.tex
\section{Related Works}\label{sec:related_works}

%%%%%%%%%%%%%%%%%%%%%%%%%%%%%%%%%%%%%%%%%%%%%%%%%%%%%%%%%%%%
%%%%%%%%%%%%%%%%%%%%%%%%%%%%%%%%%%%%%%%%%%%%%%%%%%%%%%%%%%%%

\subsection{Forecasting AF Episodes}\label{ssec:forecasting_future_af}
Early studies predicting impending AF episodes mainly focused on heart rate variability (HRV) features.
\citet{boon2016paroxysmal} demonstrated that 15-minute HRV data can effectively predict impending AF episodes, achieving an accuracy of 79.3\%.
\citet{ebrahimzadeh2018prediction} expanded this approach using a comprehensive HRV feature set and a ``mixture of experts'' classifier, achieving a robust predictive performance with an accuracy of 98.2\%.

Recent research has leveraged deep learning architectures to forecast AF onset by processing various data representations, ranging from heart rate variability and morphological features to raw ECG signals.
\citet{gilon2020forecast} proposed a deep neural network combining convolutional layers and bidirectional GRUs, trained on Holter RR interval data, and achieved an AUROC of 0.74 using a 300-beat (approximately five-minute-long) window immediately preceding the onset of AF.
\citet{tzou2021paroxysmal} extracted multi-scale wavelet features from P-wave morphology and used them to train a lightweight CNN, achieving an AUROC of 0.89 on non-AF ECGs.
\citet{gavidia2024early} developed an early warning algorithm for impending AF onset by proposing a CNN model that processed 30-second sliding RR-interval windows, validated on Holter data. The model issued alerts an average of 30.8 minutes before AF onset, achieving 83\% accuracy and an AUROC of 0.90.
The authors reported a slight performance gain when training the CNN on original ECG data compared to RRI signals.

Despite these advances, most existing studies employ generalized models trained on specific populations without addressing inter-patient variability in ECG. Such variability can limit the generalizability of ECG-based AI models and lead to inconsistent performance across patients~\citep{ding2024deep, ding2025advances}. Therefore, generalized models may fail to leverage patient-specific ECG features that are critical for accurate prediction. Indeed, \citet{jeong2024enhancing} have shown that inter-patient variability in ECG impaired deep learning performance for arrhythmia classification.

To address this gap, we introduce and evaluate a personalized fine-tuning framework for short-term forecasting of AF onset. By adapting a global model with patient-specific ECGs, we systematically assess improvements in forecasting accuracy as well as potential risks of overfitting across three large single-lead ECG databases.

%%%%%%%%%%%%%%%%%%%%%%%%%%%%%%%%%%%%%%%%%%%%%%%%%%%%%%%%%%%%
%%%%%%%%%%%%%%%%%%%%%%%%%%%%%%%%%%%%%%%%%%%%%%%%%%%%%%%%%%%%

\subsection{Personalization of Deep Learning Models in ECG Analysis}
Personalization in deep learning involves adapting models to individual users using their unique data. This approach has been applied across various domains, including recommender systems~\citep{zhang2020explainable}, personal assistants~\citep{sarikaya2017technology}, and medical applications~\citep{ding2024deep}, where user-specific preferences, behaviors, and characteristics vary significantly. In the medical field, the high variability of biosignals like blood pressures and ECGs may require personalized modeling to improve AI model performance.

For instance, \citet{hong2021deep} introduced a deep-learning model that incorporated individualized fine-tuning for dynamic and beat-to-beat blood pressure estimation. Their approach reduced mean absolute errors by adapting to patient-specific patterns. \citet{hu2023personalized} proposed a hybrid transformer model for personalized transfer learning in single-lead ECG-based sleep apnea detection. They analyzed the effects of different deep-learning model structures, label mapping lengths, and fine-tuning strategies on model performance. \citet{ng2023few} developed a personalized AF detector using a Siamese network architecture. They applied few-shot learning techniques to address the challenge of limited labeled ECG data during fine-tuning, enabling effective personalization with minimal ECG segments from a new patient.

Unlike previous studies, our research specifically targets the unique opportunities presented by ambulatory ECG data collected from wearable devices. \citet{choi2024machine} demonstrated that training serial 12-lead ECGs of a given patient can improve AF forecasting performance. A similar strategy is applicable for ambulatory ECG data, which typically provide abundant data collected over several days for each patient, enabling repetitive training and making personalized models particularly advantageous. Short-term AF forecasting, in particular, benefits substantially from personalization due to significant inter-patient variability in ECG patterns preceding AF episodes. Our study is distinct in systematically evaluating how personalized fine-tuning impacts predictive performance in this context, offering concrete evidence of improved forecasting accuracy and clinical utility.

%% file: TEX/30.methods.tex
\section{Methods}\label{sec:methods}

\subsection{Database}\label{ssec:database}
In the experiments, we utilized two public ECG databases, ICENTIA11K~\citep{tan2021icentia11k, tan2022physionet} and IRIDIA-AF~\citep{gilon2023iridia}, as well as one private ECG database, MobiCARE. These three databases contain ambulatory ECG recordings recorded by wearable devices, and they include annotations indicating timestamps of AF occurrences. The characteristics of the three databases are summarized in Table~\ref{tab:database_characteristics}.

\input{tables/database_characteristics}

ICENTIA11K~\citep{tan2021icentia11k, tan2022physionet} comprises single-lead ECG recordings collected from 11,000 patients, with monitoring periods ranging from 3 to 14 days per patient. The recordings were acquired at a sampling rate of 250 Hz using the CardioSTAT\textsuperscript{TM} in a modified lead I position. For each patient, the recordings are divided into 70-minute ECG segments, with 20 to 50 segments randomly sampled and made publicly available. Consequently, the total publicly available length for each patient ranges from approximately 1 to 2.5 days.

IRIDIA-AF~\citep{gilon2023iridia} is an ECG database consisting of 167 recordings from patients with paroxysmal AF. These recordings were obtained from 152 patients, with some patients having multiple recordings taken at distinct time periods. The recordings range in duration from 1 to 4 continuous days and were acquired at a sampling rate of 200 Hz. They were collected using the Microport Spiderview Holter recorder in lead I and II positions. To simulate single-lead ambulatory ECG recordings, we used lead I in the experiments.

MobiCARE is a private single-lead ECG dataset comprising 422 recordings prospectively collected by a single tertiary medical center (Seoul National University Hospital, Seoul, Republic of Korea) between October 2020 and November 2023. Data collection was approved by the Seoul National University Hospital Institutional Review Board and adhered to the Declaration of Helsinki revised in 2013 (IRB No: 2006-224-1138). The recordings were acquired using the mobiCARE\textsuperscript{TM} device from SEERS Technology Co., Ltd. (Seongnam-si, Republic of Korea) at a 256 Hz sampling rate, targeting patients with paroxysmal AF or other types of cardiac arrhythmias~\citep{kwon2021validation, kwon2022comparison, ahn2024three}. Each recording, captured in a modified lead II configuration, spans either 3 or 7 days. Although MobiCARE uses a different lead configuration from ICENTIA11K (modified lead I), we included it to evaluate the generalizability of the AF forecasting model across devices and to assess the effectiveness of personalization under varying lead conditions. Because the data were collected for a clinical study, the data used in our study were anonymized and not publicly available to protect patient privacy.

Our experiments utilized the AF annotations provided by the original investigators of each database.
Annotation criteria may differ across the three databases; specifically, minimum duration requirements were not explicitly documented for all sources.
We observed that the distribution of annotated AF durations varies by database, suggesting inconsistencies in the underlying annotation criteria.
Further analysis of these duration distributions is provided in \ref{sec:af_annotation_lengths}.

Notably, multiple recordings may have been sampled from the same patient in the ECG databases. However, except for the MobiCARE dataset, the mapping between recordings and patients is unavailable, making it impossible to identify multiple recordings from the same patient. In this study, we treated each recording as a unique patient.
Since our study focused on personalization within the context of a single, continuous recording, the presence of multiple recordings from the same individual did not compromise the validity of the experimental results.
In the following sections, we refer to each recording as an individual patient.

%%%%%%%%%%%%%%%%%%%%%%%%%%%%%%%%%%%%%%%%%%%%%%%%%%%%%%%%%%%%
%%%%%%%%%%%%%%%%%%%%%%%%%%%%%%%%%%%%%%%%%%%%%%%%%%%%%%%%%%%%

\subsection{Task Definition and Data Segmentation}\label{ssec:task_definition}
This study focused on a forecasting horizon of five minutes, determining whether AF onset would occur within that time frame using 60-second segments.
Similar to previous studies on short-term AF forecasting~\citep{ebrahimzadeh2018prediction, gilon2020forecast, tzou2021paroxysmal}, we defined two types of 5-minute ECG segments---referred as \textit{episodes}---based on their proximity to the AF onset.

\begin{figure}[htbp]
\centering
\includegraphics[width=\textwidth]{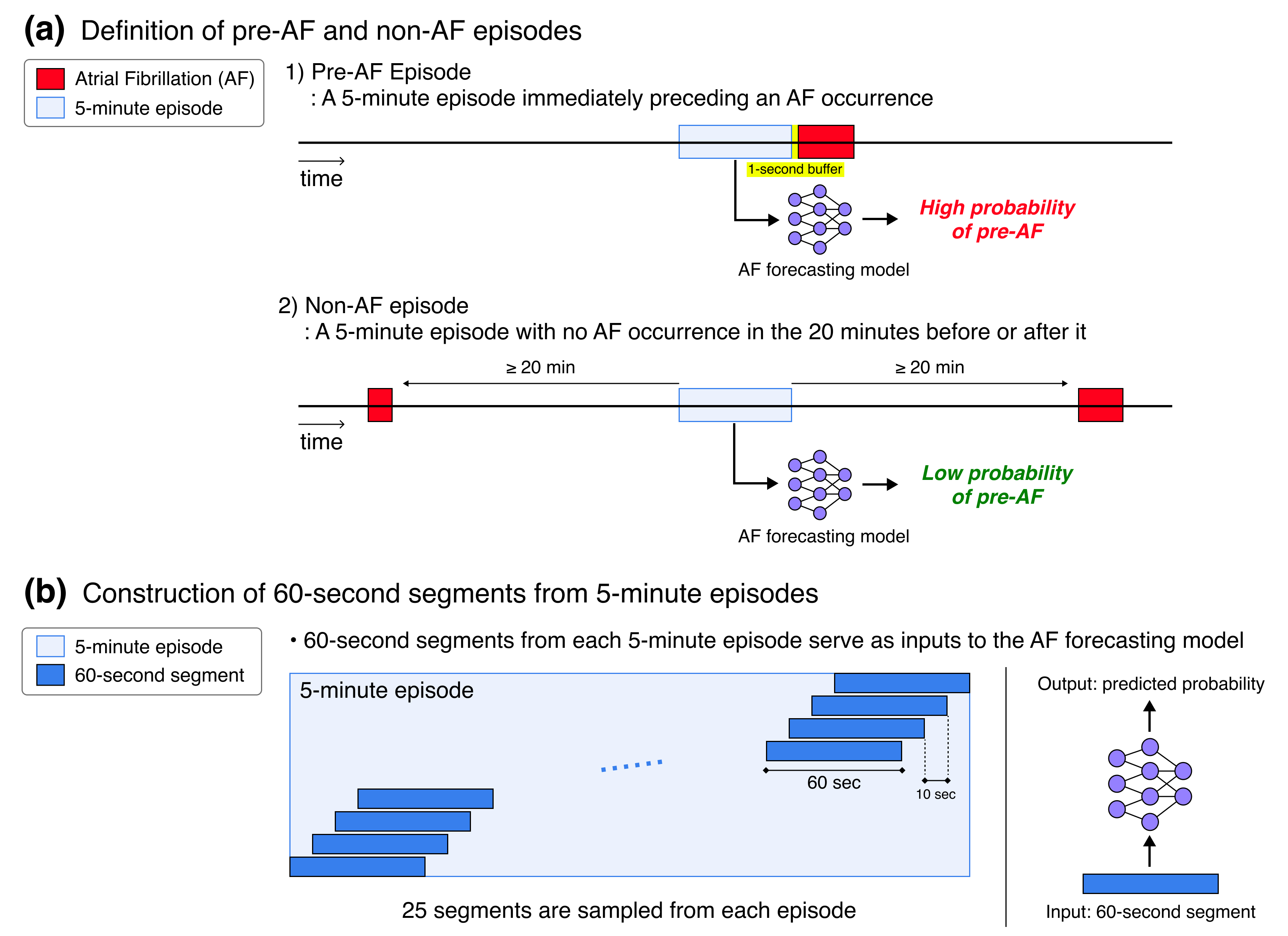}
\caption{
Data labeling and segmentation process for AF forecasting. 
\textbf{(a)} Criteria for defining \textit{pre-AF episodes} and \textit{non-AF episodes} in continuous ECG recordings.  
\textbf{(b)} Each 5-minute episode was divided into overlapping 60-second segments with a 10-second stride, yielding 25 segments per episode, which served as inputs to the forecasting model.
}
\label{fig:figure3-2-1}
\end{figure}

A \textbf{pre-AF episode} is a 5-minute segment that immediately precedes the AF onset, with a 1-second buffer to ensure no overlap with the AF. When two AFs occurred within five minutes of each other, we discarded the interval between them so that pre-AF episodes always followed a sustained period of non-AF rhythm, which is more clinically relevant for forecasting.
A \textbf{non-AF episode}, in contrast, is a 5-minute segment that is at least 20 minutes away from the onset or termination of any AF. The 20-minute margin reflects the context limitations in the ICENTIA11K dataset, where each continuous recording only spans 70 minutes. Figure~\ref{fig:figure3-2-1}a illustrates the selection criteria for pre-AF and non-AF episodes.

Each 5-minute episode was segmented into overlapping 60-second windows with a 10-second stride, resulting in 25 segments per episode. These segments served as input to the AF forecasting model. Figure~\ref{fig:figure3-2-1}b illustrates this segmentation process. The task was defined as a binary classification problem: determining whether a given 60-second segment originated from a pre-AF or a non-AF episode. The model produced an output between 0 and 1, where values closer to 1 indicated a higher probability that the segment belonged to a pre-AF episode (i.e., AF onset within the next 5 minutes), and values closer to 0 indicated a higher probability that the segment was from the non-AF episode. Because all 25 segments were systematically sampled from the same 5-minute episode, this design enabled consistent analysis of model performance by temporal distance from AF onset.

%%%%%%%%%%%%%%%%%%%%%%%%%%%%%%%%%%%%%%%%%%%%%%%%%%%%%%%%%%%%
%%%%%%%%%%%%%%%%%%%%%%%%%%%%%%%%%%%%%%%%%%%%%%%%%%%%%%%%%%%%

\subsection{Experimental Design}\label{ssec:experimental_design}

\begin{figure}[htbp]
\centering
\includegraphics[width=\textwidth]{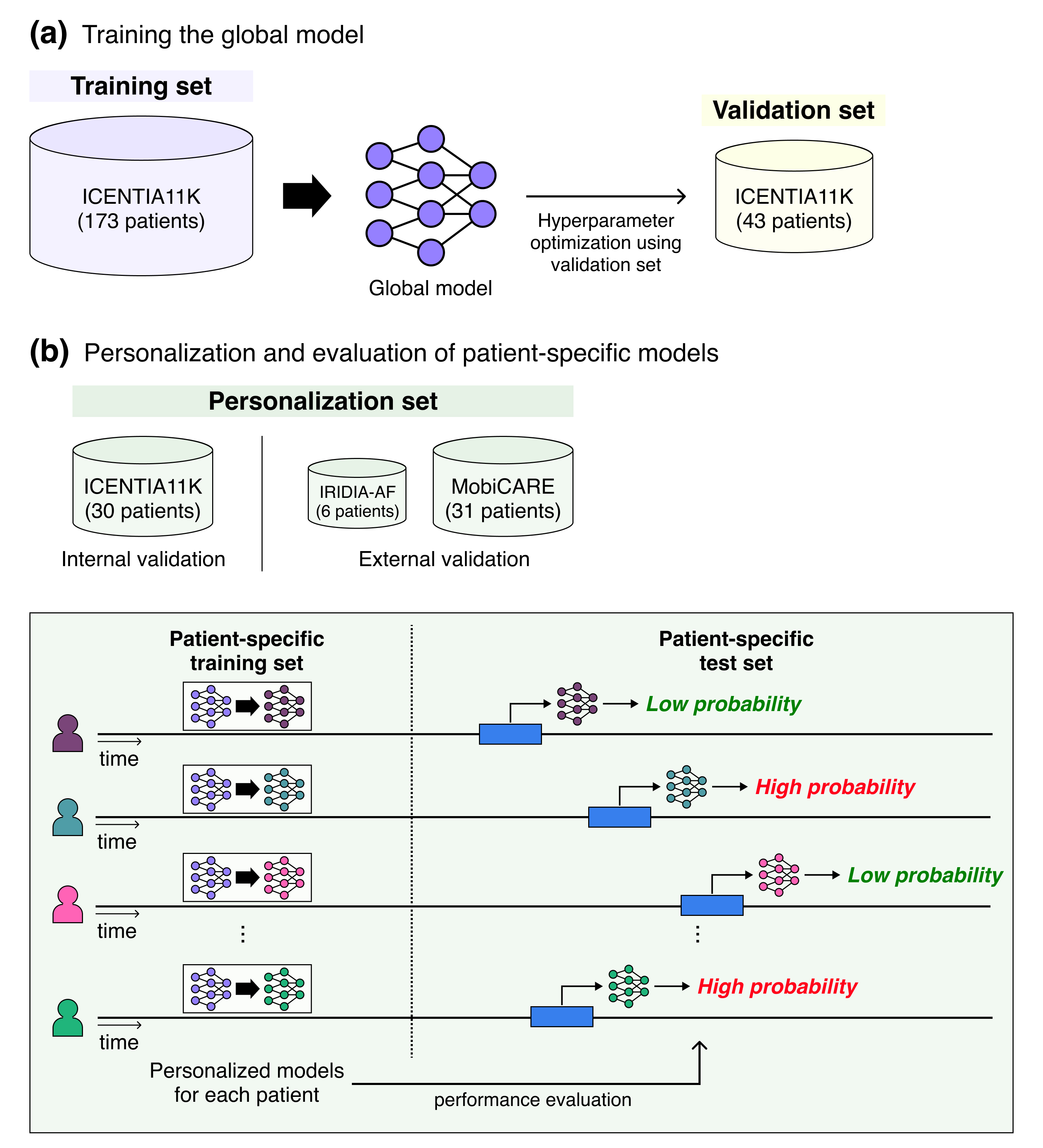}
\caption{
Overview of the experimental design. 
\textbf{(a)} Training the global model. A deep learning model was trained on the ICENTIA11K dataset to capture generalizable patterns preceding imminent AF.
\textbf{(b)} Personalization and evaluation of patient-specific models. The global model was fine-tuned using a small amount of labeled data from individual patients (patient-specific training set) and evaluated on separate, later segments from the same patients (patient-specific test set), enabling assessment of personalization effectiveness.
}
\label{fig:figure3-3-1}
\end{figure}

The goal of our experiments was to evaluate whether the personalized model outperformed the global model in short-term forecasting of AF onset, as described in Section~\ref{ssec:task_definition}. To this end, we organized the experiments into two stages: (1) training a global model and (2) training and evaluating personalized models. An overview of the experimental design is presented in Figure~\ref{fig:figure3-3-1}.

In the first stage, we trained a \textbf{global model} using a pooled dataset of ECG recordings from multiple patients, simulating a development scenario where a model was constructed from large-scale, pre-existing data. This model was intended to capture generalizable features of imminent AF that apply across individuals. We used the ICENTIA11K dataset for this stage due to its relatively large patient population.
The data were divided into a \textit{training set} and a \textit{validation set}, with the latter used for hyperparameter tuning and early stopping. 
The resulting global model served as a baseline for evaluating its performance against personalized models.

In the second stage, we assessed the effect of personalization by fine-tuning the global model with a small amount of labeled data from each target patient, thereby obtaining a \textbf{personalized model} for each individual. To this end, we defined a \textit{personalization set}, consisting of selected patients from the ICENTIA11K, IRIDIA-AF, and MobiCARE datasets. The detailed criteria for selecting patients for the personalization sets are provided in Section~\ref{ssec:patient_selection_data_preparation}.

Each patient’s recording in the personalization set was temporally split into two spans: the earlier span was used for fine-tuning the global model (referred to as the \textit{patient-specific training set}), and the later span is used to evaluate the resulting personalized model (the \textit{patient-specific test set}).
For ICENTIA11K, the patient-specific training set comprised the first 20 segments (approximately 24 hours), while for IRIDIA-AF and MobiCARE, the first 24 hours of each recording were used. The remaining spans, used as patient-specific test sets, ranged from one to six days depending on the length of each patient’s recording.

%%%%%%%%%%%%%%%%%%%%%%%%%%%%%%%%%%%%%%%%%%%%%%%%%%%%%%%%%%%%
%%%%%%%%%%%%%%%%%%%%%%%%%%%%%%%%%%%%%%%%%%%%%%%%%%%%%%%%%%%%

\subsection{Patient Selection and Data Preparation}\label{ssec:patient_selection_data_preparation}
\begin{figure}[htbp]
\centering
\includegraphics[width=\textwidth]{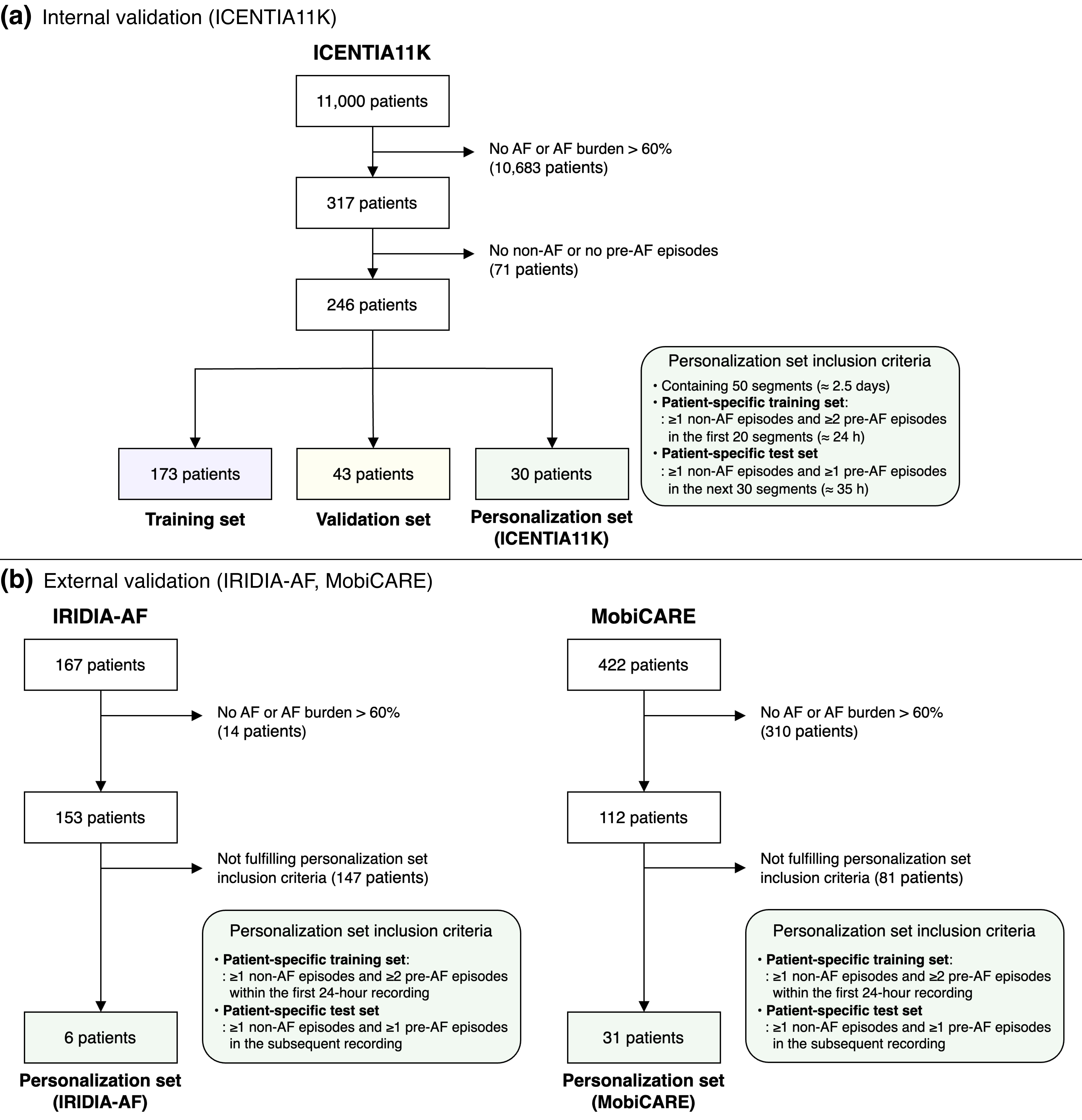}
\caption{Patient inclusion algorithm.
\textbf{(a)} Patients in the ICENTIA11K dataset were divided into training, validation, and personalization sets for internal validation.
\textbf{(b)} Patients in the IRIDIA-AF and MobiCARE datasets were selected to form personalization sets for external validation.
}
\label{fig:patient_inclusion_algorithm}
\end{figure}

The overview of the patient inclusion algorithm for our experiments is presented in Figure~\ref{fig:patient_inclusion_algorithm}. To focus on clinically relevant scenarios---providing alerts to patients experiencing intermittent AF occurrences---we excluded patients without any AF episode or those with an AF burden exceeding 60\%. Furthermore, we excluded patients that contain neither non-AF nor pre-AF episodes from the experiments.
This ensured the model learned within-patient differences between pre-AF and distant-from-AF rhythms, rather than separating high- from low-burden patients---an aspect unrelated to patient-specific forecasting.

Since the ICENTIA11K dataset contained the largest number of samples, we used it for internal validation by splitting it into training, validation, and personalization sets. The training and validation sets were used to train the global model, while the personalization set was used to evaluate the effect of personalization.
To ensure a fair evaluation, all patient data used in the personalization sets were strictly excluded from the global model's training and validation sets.
IRIDIA-AF and MobiCARE datasets were used for external validation, where personalization sets were constructed to assess the trained model's performance and the effectiveness of personalization when applied to different data sources.

% In the recordings of the personalization set, the patient-specific training set consisted of the first 24 hours, while the patient-specific test set included the remaining portion of the recording.
To fine-tune the global model using the patient-specific training set and evaluate its performance with the patient-specific test set, both sets were required to contain pre-AF and non-AF episodes. Thus, the following inclusion criteria were applied to the personalization set:
\begin{itemize}
    \item[1)] The patient-specific training set (first 24 hours) contained at least one non-AF episode and at least two pre-AF episodes.  
    \item[2)] The patient-specific test set (remaining portion) contained at least one non-AF episode and at least one pre-AF episode.
\end{itemize}

For the ICENTIA11K dataset, 246 patients were identified as having an AF burden of less than 60\% and containing both non-AF and pre-AF episodes. Recordings in ICENTIA11K were divided into 70-minute segments, with a maximum of 50 segments, and only patients with exactly 50 segments were included in the personalization set. We randomly sampled 30 patients that met the inclusion criteria for the personalization set and split the remaining patients into an 8:2 ratio to form the training and validation sets. This resulted in 173 patients for training, 43 for validation, and 30 for personalization in the ICENTIA11K dataset.

For the IRIDIA-AF and MobiCARE datasets, which were used as external validation sets, patients that met the inclusion criteria for the personalization set were selected. As a result, six patients were chosen from the IRIDIA-AF dataset and 31 from the MobiCARE dataset. Table~\ref{tab:characteristics_of_selected_samples} represents the age and sex distributions for patients corresponding to the patients included in the experiments.

\input{tables/characteristics_of_selected_samples}

The typical recordings in our experimental datasets contained significantly more non-AF episodes than pre-AF episodes, and the number of episodes extracted from each patient varied greatly. To balance the dataset and prevent episodes from a few patients from dominating the data, we capped the number of pre-AF episodes per patient at 10 and limited the number of non-AF episodes to twice the number of pre-AF episodes per patient. Overall, the total number of episodes per patient in each set was capped at 30, with a class ratio of approximately 1:2 for pre-AF and non-AF in most patients.

As explained in Section~\ref{ssec:task_definition}, we sampled 25 60-second segments from each episode as inputs to the deep learning model, resulting in up to 750 segments per patient in each set. The numbers of 5-minute episodes and 60-second segments for each set are detailed in Table~\ref{tab:num_episodes_training_validation} and \ref{tab:num_episodes_personalization}.

%%%%%%%%%%%%%%%%%%%%%%%%%%%%%%%%%%%%%%%%%%%%%%%%%%%%%%%%%%%%
%%%%%%%%%%%%%%%%%%%%%%%%%%%%%%%%%%%%%%%%%%%%%%%%%%%%%%%%%%%%

\subsection{Preprocessing}\label{ssec:preprocessing}
Robust signal processing is essential for forecasting from noisy ambulatory recordings; we therefore implemented a quality-control and filtering pipeline before model input.

The samples from ICENTIA11K, used for model training and internal validation, were recorded at a sampling rate of 250 Hz. To maintain consistency across datasets, recordings from IRIDIA-AF and MobiCARE were resampled to 250 Hz. For IRIDIA-AF, lead I recordings were selected to match the recorded lead in ICENTIA11K.

ECG recordings from wearable devices often contain significant noise because they are collected while patients are active. 
To construct the dataset from continuous multi-day recordings, we implemented a preprocessing pipeline to filter noisy data and ensure signal quality. Long-term recordings were divided into non-overlapping 5-minute episodes. For each recording, an episode was randomly sampled and subjected to a signal quality test. We identified `flat' signals--typically indicative of ECG patch detachment--where the amplitude remained constant for over 1 second. Any episode containing more than 30 seconds of cumulative flat signal was discarded.
The surviving episodes were zero-mean normalized, and a first-order Butterworth high-pass filter with a 0.5 Hz cutoff frequency was applied to mitigate baseline wander. If the absolute amplitude of an episode remained below a threshold of 0.03 after filtering, it was considered to lack meaningful physiological information and was discarded. Otherwise, it was included in the patient’s dataset. This sampling process continued until a predefined cap on the number of episodes was reached. Finally, the selected 5-minute episodes were partitioned into 60-second segments using a 10-second stride. Each segment was re-normalized to zero mean and high-pass filtered at 0.5 Hz prior to model processing.
%%%%%%%%%%%%%%%%%%%%%%%%%%%%%%%%%%%%%%%%%%%%%%%%%%%%%%%%%%%%
%%%%%%%%%%%%%%%%%%%%%%%%%%%%%%%%%%%%%%%%%%%%%%%%%%%%%%%%%%%%

\subsection{Data Augmentation}\label{ssec:data_augmentation}
\begin{figure}[htbp]
\centering
\includegraphics[width=\textwidth]{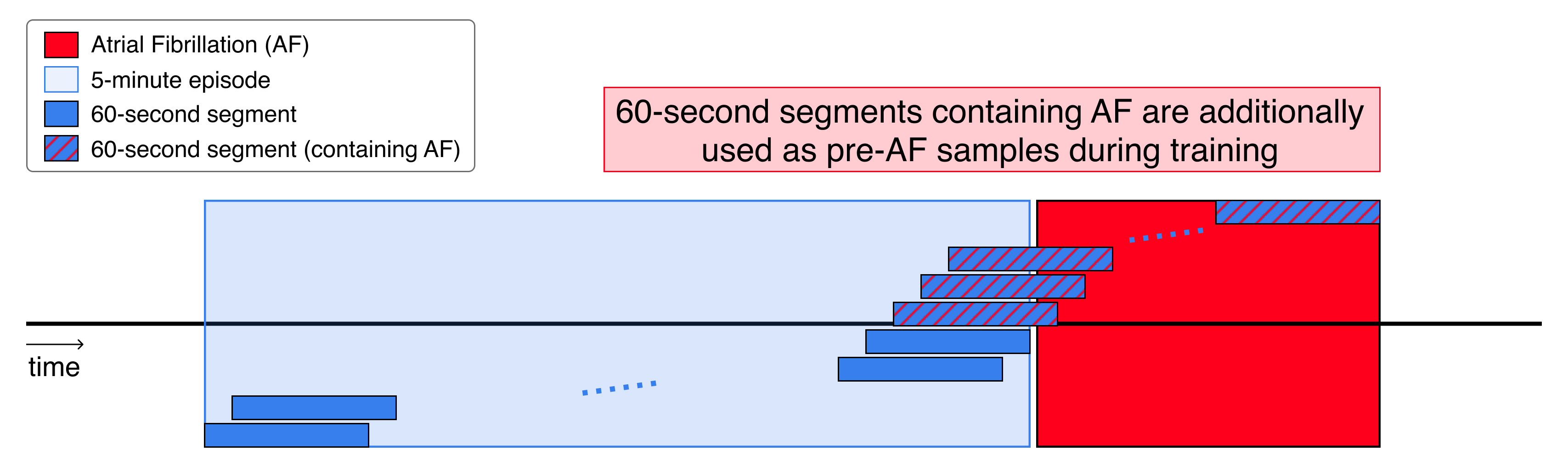}
\caption{
Illustration of how 60-second segments containing AF were additionally included as pre-AF samples during training.
}
\label{fig:figure3-6-1}
\end{figure}

To improve the training process, we applied a data augmentation scheme by scaling each 60-second segment input randomly between 70\% and 130\%~\citep{suh2021learning}. To further enhance learning with limited data, we incorporated 60-second segments containing AF as pre-AF samples during training only. 
This augmentation supplies additional positive-class examples whose morphology is contiguous with imminent onset; following \citet{gavidia2024early}, who defined AF risk as the summed probability of pre-AF and AF, we expected such inputs to yield high outputs.
For each pre-AF episode, we generated additional 60-second segments by shifting 10 seconds toward the AF onset, starting from the endpoint of the pre-AF episode. This process continued until the termination of AF or for 5 minutes after the first AF-containing segment. Figure~\ref{fig:figure3-6-1} illustrates this procedure.

%%%%%%%%%%%%%%%%%%%%%%%%%%%%%%%%%%%%%%%%%%%%%%%%%%%%%%%%%%%%
%%%%%%%%%%%%%%%%%%%%%%%%%%%%%%%%%%%%%%%%%%%%%%%%%%%%%%%%%%%%

\subsection{Model Architecture and Performance Evaluation}\label{ssec:model_architecture_and_performance_evaluation}
We employed a deep learning model designed for ECG data, based on the ResNet~\citep{he2016deep} architecture. Specifically, we developed a modified 18-layer ResNet with 1x15 sized kernels to process 1D ECG signals (Figure~\ref{fig:model_architecture}). The \ref{sec:implementation_details} provides details of the hyperparameters and the search procedure used to optimize them.

Seven metrics were used to assess the performance of the deep-learning models: area under the receiver operating characteristic curve (AUROC), area under the precision-recall curve (AUPRC), sensitivity, specificity, positive predictive value (PPV), negative predictive value (NPV), and F1-score. To compute metrics requiring binary predictions---sensitivity, specificity, PPV, and NPV---we determined patient-specific thresholds that maximized the Youden index~\citep{youden1950index} on the patient-specific test set. The threshold-dependent metrics represent values at the optimal operating point and are intended for descriptive comparison between models rather than as estimates of prospectively achievable performance. The performance for each patient was obtained by averaging the results of five training and evaluation runs with different random seeds and initializations.
Final metrics were reported as the average across all patients in the personalization set, with 95\% confidence intervals computed based on the number of patients in each dataset ($n=30$ for ICENTIA11K, $n=6$ for IRIDIA-AF, and $n=31$ for MobiCARE). Statistical significance between global and personalized models was assessed using a two-sided paired $t$-test.

%% file: tables/database_characteristics.tex
\begin{table}[htb]
  \centering
  \caption{Characteristics of the ECG databases.}
  \label{tab:database_characteristics}

  \begin{threeparttable}
  % X = ragged-right, width automatically apportioned
  \begin{tabularx}{\textwidth}{@{}l *{3}{>{\raggedright\arraybackslash}X}@{}}
    \toprule
                     & \textbf{ICENTIA11K} & \textbf{IRIDIA-AF} & \textbf{MobiCARE} \\
    \midrule
    Sampling rate    & 250 Hz      & 200 Hz      & 256 Hz \\
    Patients         & 11\,000      & 152         & 379   \\
    Recordings\tnote{a}  & 11\,000      & 167         & 422   \\
    Leads            & Lead I      & Lead I, Lead II\tnote{b} & Lead II \\
    Record duration  & 1–2.5 days (70-min segments) & 1–4 days & 3–7 days \\
    \bottomrule
  \end{tabularx}

  \begin{tablenotes}[flushleft]\footnotesize
    \item[a] We treat each unique continuous recording as a distinct patient.
    \item[b] Only Lead I was analyzed.
  \end{tablenotes}
  \end{threeparttable}
\end{table}

%% file: tables/characteristics_of_selected_samples.tex
\begin{table}[htbp]
  \centering
  \caption{Characteristics of the selected patients.}
  \label{tab:characteristics_of_selected_samples}

  \begin{threeparttable}
    \footnotesize          % keeps the table compact; remove if unnecessary
    % Layout: first column = left-aligned text,
    %          next four columns = centred (or ragged-left/right as desired)
    \begin{tabularx}{\textwidth}{@{}l>{\raggedright\arraybackslash}X
                                 *{3}{>{\centering\arraybackslash}X}@{}}
      \toprule
      \multicolumn{2}{l}{} &
      \textbf{Internal validation} &
      \multicolumn{2}{c}{\textbf{External validation}} \\ 
      \cmidrule(lr){3-3}\cmidrule(l){4-5}
      \multicolumn{2}{l}{} &
      \textbf{ICENTIA11K} & \textbf{IRIDIA-AF} & \textbf{MobiCARE} \\
      \midrule
      \multicolumn{2}{l}{Patients\tnote{a}} &
        246\,(\textit{train}/\textit{val}/\textit{pers}) & 6\,(\textit{pers}) & 31\,(\textit{pers}) \\
      \multicolumn{2}{l}{Age, years (mean ± SD)} &
        62.2 ± 17.4\tnote{b} & 68.2 ± 9.7 & 58.1 ± 10.4 \\
      \multirow{3}{*}{Sex, \%} & Male &
        42.6\tnote{b} & 66.6 & 77.4 \\
      & Female &
        45.3\tnote{b} & 33.3 & 22.6 \\
      & Unknown &
        12.2\tnote{b} & 0.0  & 0.0 \\
      \bottomrule
    \end{tabularx}

    \begin{tablenotes}[flushleft]\footnotesize
      \item[a] \textit{train} = training set; \textit{val} = validation set; \textit{pers} = personalization set.
      \item[b] Statistics taken from the original ICENTIA11K publication;
               individual-level demographics are not provided in the dataset.
    \end{tablenotes}
  \end{threeparttable}
\end{table}

%% file: TEX/40.results.tex
\section{Results}\label{sec:results}

\subsection{Impact of Personalized Fine-tuning on AF Forecasting performance}\label{ssec:impact_of_personalized_training}

To assess whether personalized models improved predictive performance over a global model, we evaluated both using personalization sets: ICENTIA11K for internal validation and IRIDIA-AF and MobiCARE for external validation.

\subsubsection{Internal Validation Results}\label{sssec:internal_validation_results}
\begin{figure}[htbp]
\centering
\includegraphics[width=\textwidth]{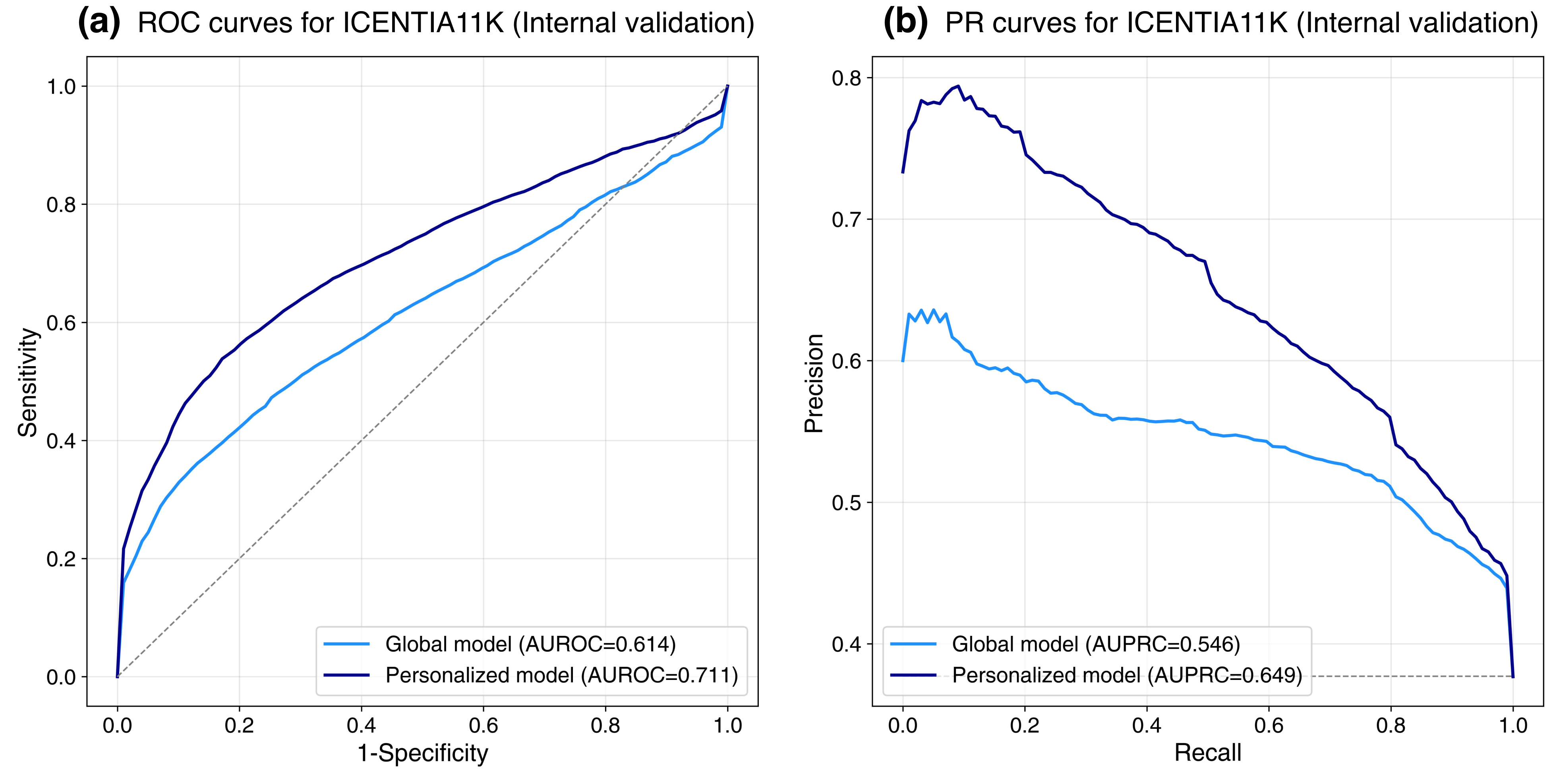}
\caption{
Comparison of global and personalized models on ICENTIA11K. (a) ROC curves. (b) PR curves.
}
\label{fig:icentia11K_results}
\end{figure}

Table~\ref{tab:internal_validation_results} compares the performance of both models using the ICENTIA11K personalization set. Overall, the personalized model outperformed the global model across all metrics, suggesting that adapting the model to individual patients improved predictive accuracy. The personalized model achieved a significantly higher AUROC (0.711 vs. 0.614, $p=0.0460$) and AUPRC (0.649 vs. 0.546, $p=0.0212$), indicating improved discrimination between AF and non-AF cases. Other metrics, including sensitivity, specificity, PPV, NPV, and F1-score showed numerical improvement but did not meet statistical significance. Figure~\ref{fig:icentia11K_results} shows a comparison of ROC curves and PR curves for both models.

\input{tables/internal_validation_results}

\clearpage

\subsubsection{External Validation Results}\label{sssec:external_validation_results}

Table~\ref{tab:external_validation_results_iridia-af} and Table~\ref{tab:external_validation_results_mobicare} compare the performance of both models on the IRIDIA-AF and MobiCARE personalization sets. Generally, the external validation results aligned with those of the internal validation.

\begin{figure}[htbp]
\centering
\includegraphics[width=\textwidth]{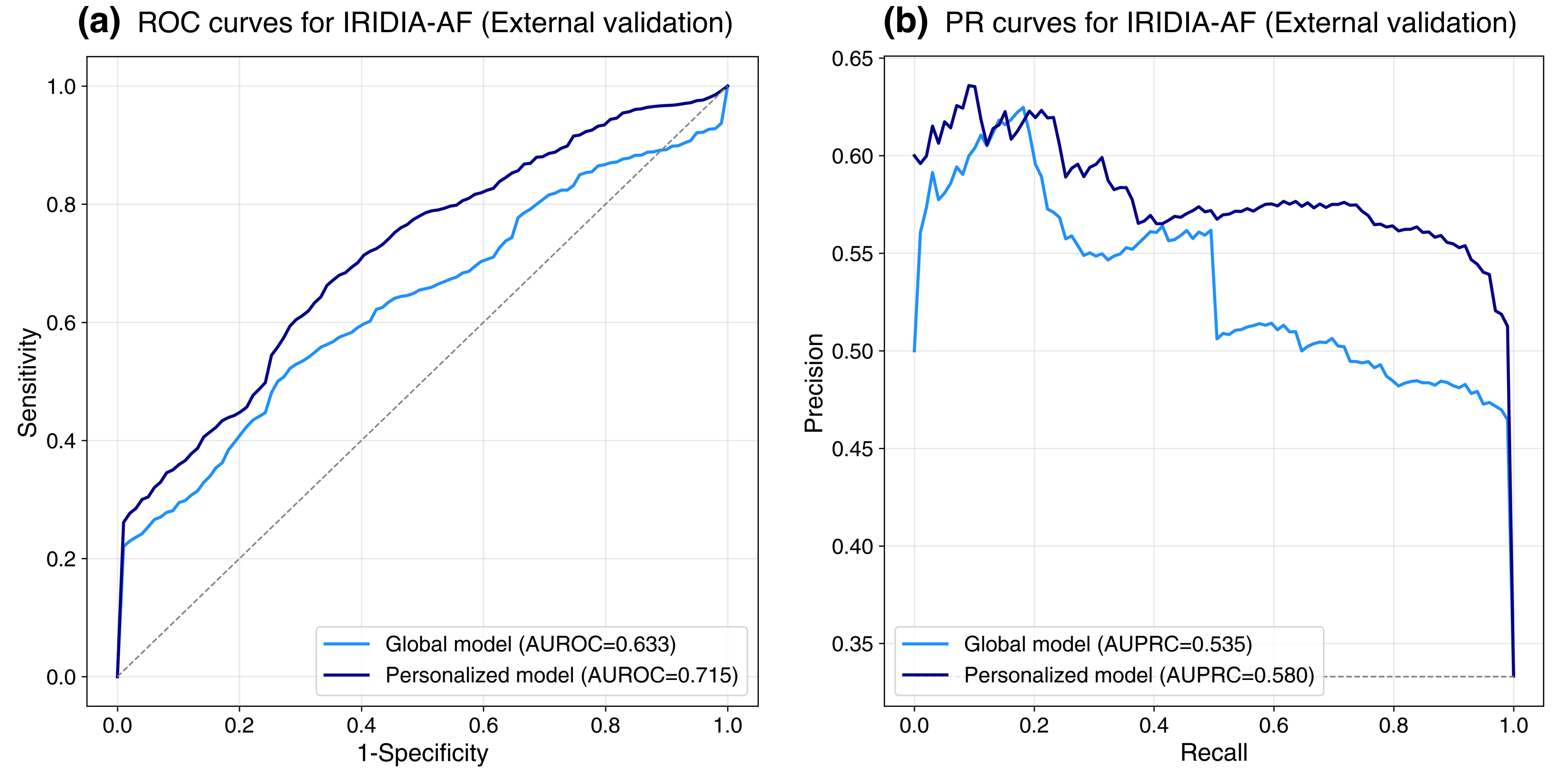}
\caption{
Comparison of global and personalized models on IRIDIA-AF. (a) ROC curves. (b) PR curves.
}
\label{fig:iridia-af_results}
\end{figure}

In the IRIDIA-AF personalization set, the personalized model achieved numerically higher AUROC (0.715 vs. 0.633, $p=0.1954$) and AUPRC (0.580 vs. 0.535, $p=0.5335$), but did not reach statistical significance (Table~\ref{tab:external_validation_results_iridia-af} and Figure~\ref{fig:iridia-af_results}). Given the small sample size of the IRIDIA-AF personalization set (i.e., only six patients), the study may have been underpowered to demonstrate statistical significance. Numerical improvement was also observed in other metrics.

\begin{figure}[htbp]
\centering
\includegraphics[width=\textwidth]{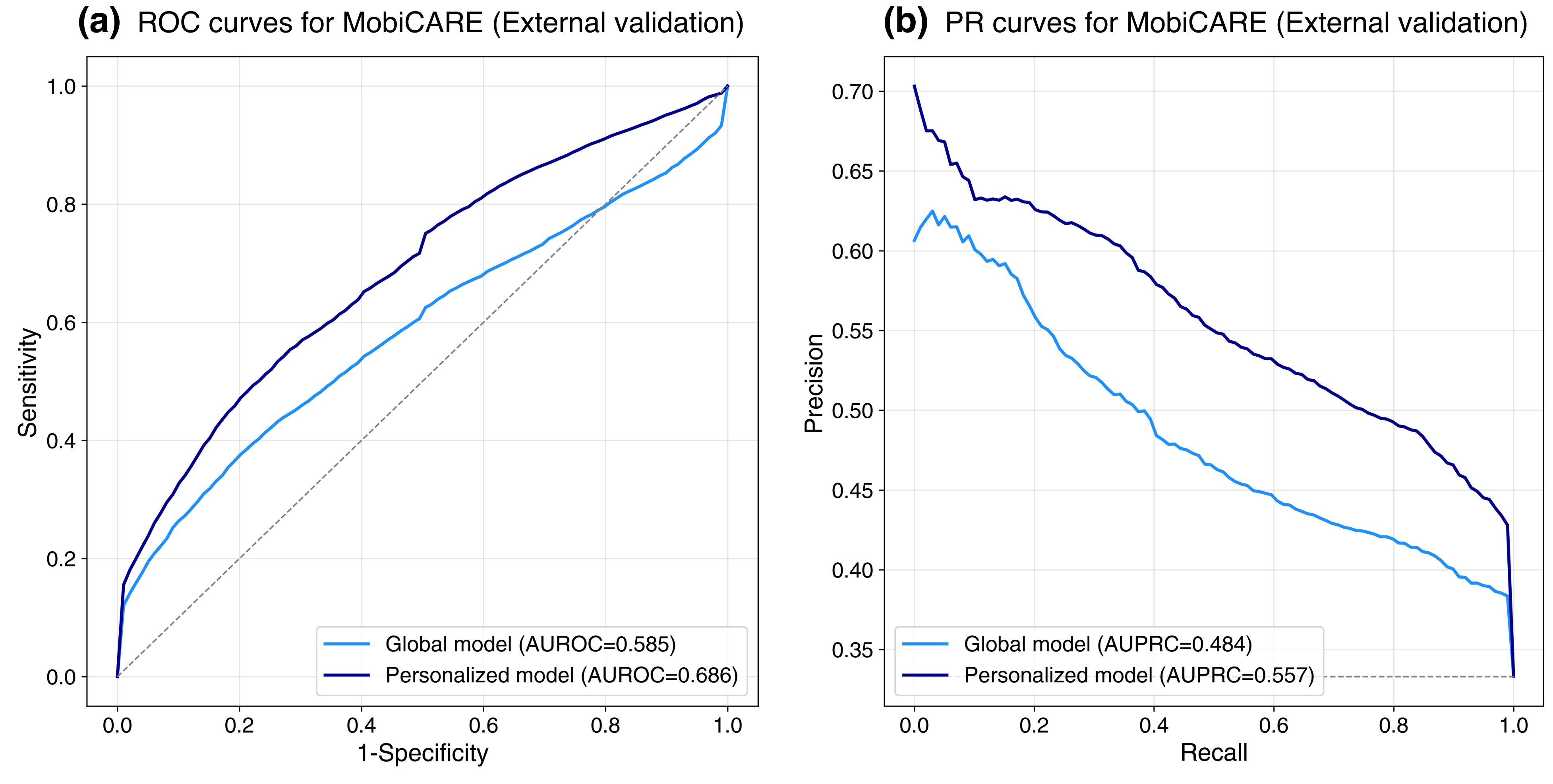}
\caption{
Comparison of global and personalized models on MobiCARE. (a) ROC curves. (b) PR curves.
}
\label{fig:mobicare_results}
\end{figure}

In the MobiCARE test set, the personalized model demonstrated significantly improved results for AUROC (0.686 vs. 0.585, $p=0.0179$) and AUPRC (0.557 vs. 0.484, $p=0.0517$) (Table~\ref{tab:external_validation_results_mobicare} and Figure~\ref{fig:mobicare_results}). Sensitivity, NPV, and F1-score also showed statistically significant improvements. The global model performed the worst on this dataset, likely due to differences in dataset characteristics: it was trained using lead I recordings, whereas the MobiCARE dataset consists of lead II recordings. These results indicate that personalization effectively adapts the AF forecasting model to datasets with different characteristics.

\input{tables/external_validation_results_iridia-af}

\input{tables/external_validation_results_mobicare}

%%%%%%%%%%%%%%%%%%%%%%%%%%%%%%%%%%%%%%%%%%%%%%%%%%%%%%%%%%%%
%%%%%%%%%%%%%%%%%%%%%%%%%%%%%%%%%%%%%%%%%%%%%%%%%%%%%%%%%%%%
\clearpage

\subsection{Influence of Training Episode Count on Personalization Effectiveness}\label{ssec:influence_of_training_episode_count}

Since we used the first 24-hour recording of each patient as the patient-specific training set, the number of episodes available for personalized training varied among patients. For some, only a few episodes were available due to the limited occurrences of AF in their training set. In such cases, the model may not have enough data to learn patient-specific patterns for forecasting impending AF, which could lead to overfitting. We hypothesized that the key factor influencing the effectiveness of personalization was the number of episodes used for training.

\begin{figure}[htbp]
\centering
\includegraphics[width=\textwidth]{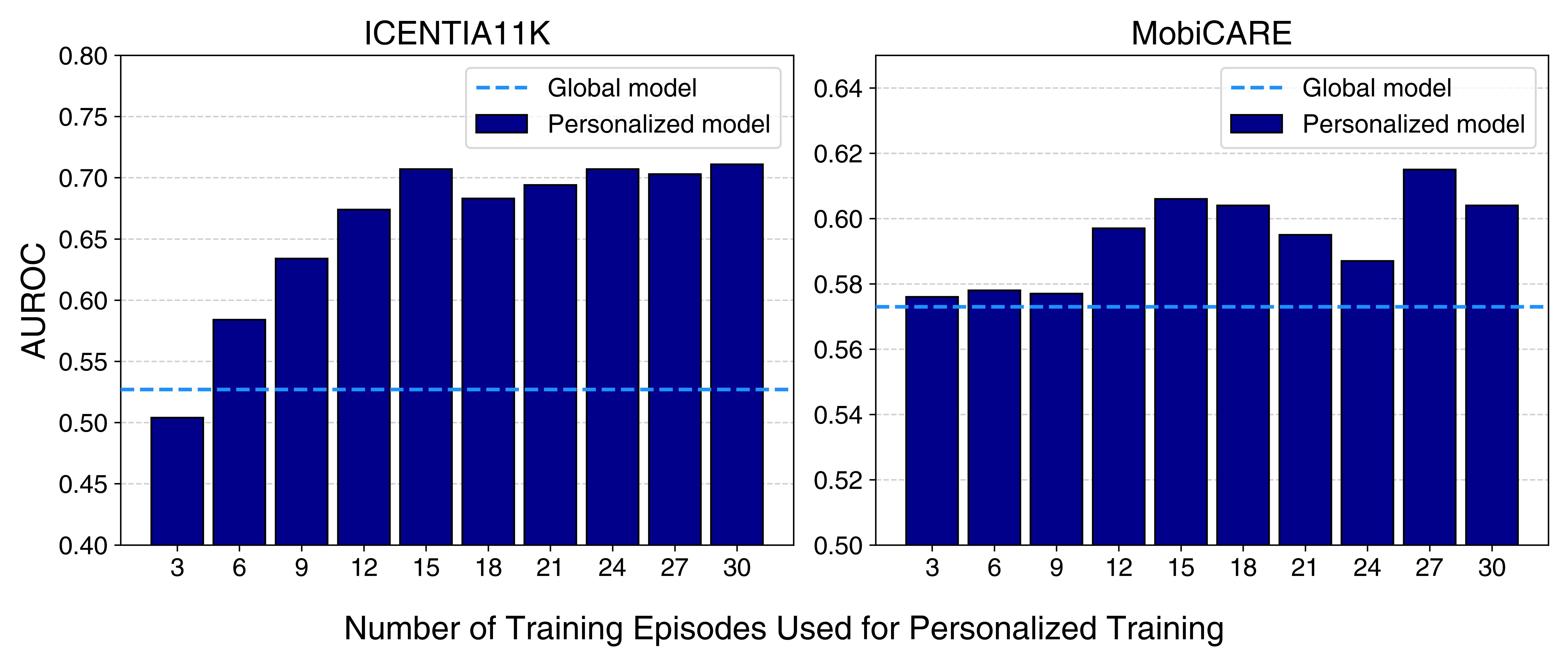}
\caption{
Effect of the number of training episodes on personalization performance. AUROC values of personalized models are shown for ICENTIA11K and MobiCARE. The dashed line indicates the performance of the global model without personalization.
}
\label{fig:effect_n_training_samples}
\end{figure}

To isolate the effect of the number of training episodes per patient from other factors, we selected patients with 30 training episodes (10 pre-AF and 20 non-AF episodes) from the ICENTIA11K and MobiCARE personalization sets. This selection included eight patients from ICENTIA11K and nine from MobiCARE. We then varied the number of training episodes through random sampling and compared the resulting AUROCs after personalization, as shown in Figure~\ref{fig:effect_n_training_samples}.

Overall, the results showed that the number of training episodes used for personalized training positively influenced forecasting performance. In both datasets, AUROC was lowest when only three training episodes (one pre-AF and two non-AF) were used, with values of 0.504 for ICENTIA11K and 0.573 for MobiCARE. In ICENTIA11K, AUROC was 0.527 without personalization and exceeded 0.7 as the number of training episodes increased. In MobiCARE, AUROC improved from 0.573 without personalization to a peak of 0.615 when 27 training episodes were used. Variability in the increasing trend arose from differences in the sampled episodes, as the specific choice of training episodes substantially affected model performance. These findings suggested that increasing the number of training episodes generally improved personalization performance.

%%%%%%%%%%%%%%%%%%%%%%%%%%%%%%%%%%%%%%%%%%%%%%%%%%%%%%%%%%%%
%%%%%%%%%%%%%%%%%%%%%%%%%%%%%%%%%%%%%%%%%%%%%%%%%%%%%%%%%%%%

\subsection{Forecasting Performance by Distance from AF Onset}\label{sec:prediction_performance_by_distance}

\begin{figure}[htbp]
\centering
\includegraphics[width=0.95\textwidth]{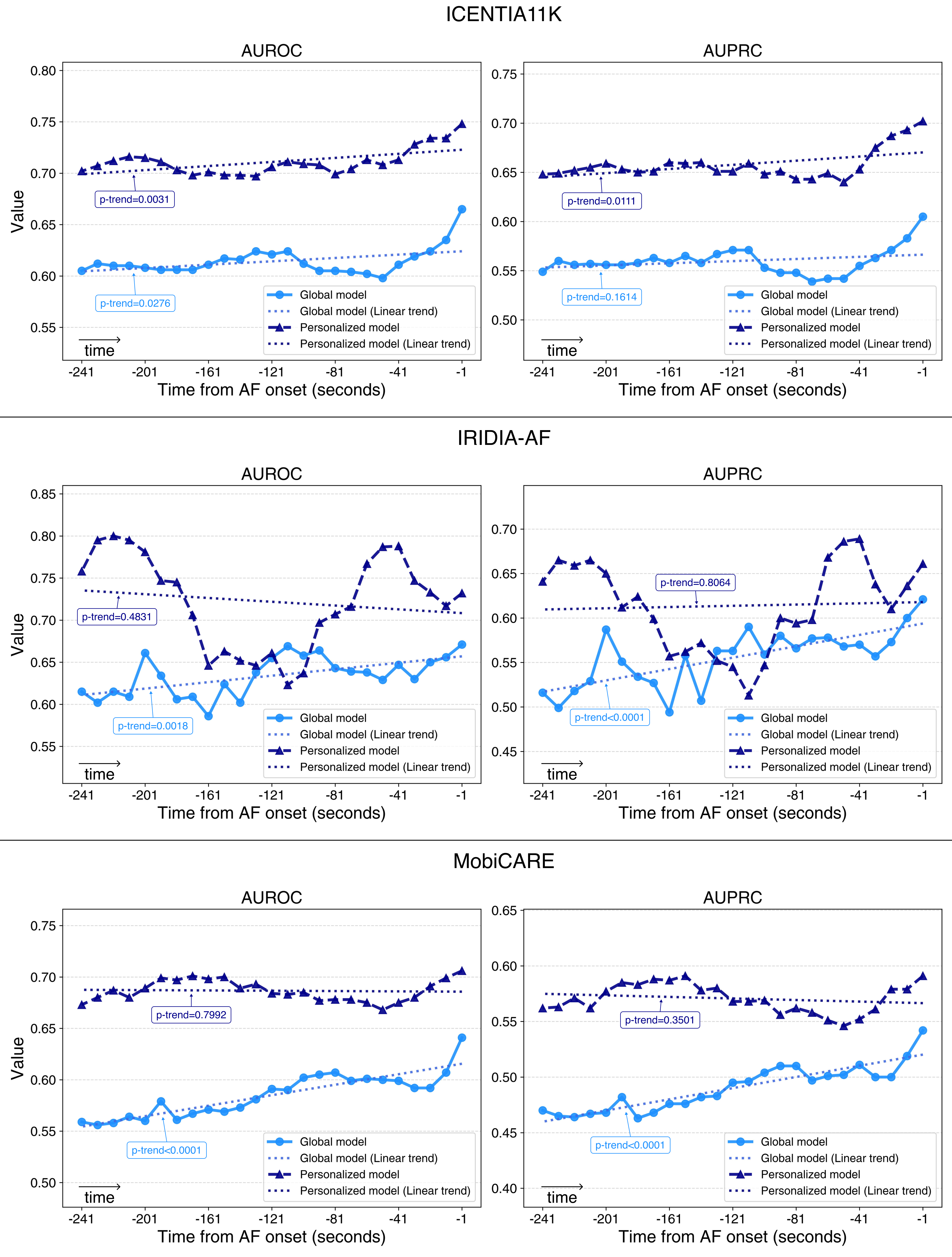}
\caption{
Forecasting performance with respect to temporal distance from AF onset. AUROC and AUPRC values are shown for global and personalized models across the ICENTIA11K, IRIDIA-AF, and MobiCARE datasets.
Performance was evaluated on 60-second segments sampled with a 10-second stride from 5-minute episodes.
A dotted line indicates the linear regression fit to the performance values over time.
}
\label{fig:window-wise_performances}
\end{figure}

As described in Section~\ref{ssec:task_definition}, we sampled 60-second segments from each 5-minute episode using a 10-second stride as input to the deep learning model. With this design, we analyzed how performance metrics vary with the distance from AF onset in pre-AF episodes. For each patient, we calculated the metrics and aggregated the results based on their distance from the onset of AF. Figure~\ref{fig:window-wise_performances} shows how prediction performance (AUROC, AUPRC, sensitivity, and specificity) changed with respect to temporal distance from AF onset.

For the global model, a significant increasing trend in AUROC was observed as the forecasting window approached AF onset across three datasets ($p$-for-trend$=0.0276$ for ICENTIA11K, $0.0018$ for IRIDIA-AF, and $<0.0001$ for MobiCARE). AUPRC values showed a similar pattern, although the increase in ICENTIA11K did not reach statistical significance ($p$-for-trend$=0.1614$ for ICENTIA11K, $<0.0001$ for IRIDIA-AF and MobiCARE).
At one second before AF onset, AUROC values were 0.665 for ICENTIA11K, 0.671 for IRIDIA-AF, and 0.641 for MobiCARE, representing the highest values across all time points. AUPRC values followed the same pattern. In IRIDIA-AF, performance fluctuations over time were more pronounced than in the other datasets, likely due to the small number of patients ($n=6$) in its personalization set, which made the results more sensitive to variance. These findings were consistent with previous studies on AF onset forecasting, which also reported a decline in performance as the forecasting window moved further from AF onset~\citep{gilon2020forecast, gavidia2024early}.

For the personalized model, ICENTIA11K exhibited temporal patterns similar to those of the global model, but with overall performance improvements. As in the global model, AUROC and AUPRC values at one second before AF onset were the highest (0.748 and 0.702, respectively), with significant increasing trends ($p$-for-trend$=0.0031$ for AUROC and $0.0111$ for AUPRC).
However, this temporal trend was not maintained after personalization in the IRIDIA-AF and MobiCARE datasets. In IRIDIA-AF, the temporal dynamics changed noticeably after personalization. AUROC and AUPRC improved markedly in certain intervals (231–211 s and 61–41 s before onset) but fluctuated over time, again reflecting the small cohort ($n=6$). In MobiCARE, the highest performance still occurred one second before AF onset (AUROC$=0.706$, AUPRC$=0.591$), but notable performance gains were also observed in segments farther from onset. This pattern suggests that, in addition to leveraging general ECG features closer to AF onset, the personalized model may have learned patient-specific characteristics that were independent of temporal distance, thereby enabling AF forecasting in a manner different from the global model. Because the global model was originally trained on ICENTIA11K, fine-tuning on datasets with different characteristics such as IRIDIA-AF and MobiCARE may have driven the personalized model to focus more heavily on patient-specific features.

%%%%%%%%%%%%%%%%%%%%%%%%%%%%%%%%%%%%%%%%%%%%%%%%%%%%%%%%%%%%
%%%%%%%%%%%%%%%%%%%%%%%%%%%%%%%%%%%%%%%%%%%%%%%%%%%%%%%%%%%%

\subsection{Indicators of Impending AF Occurrences
}\label{ssec:indicators_of_imminent_af_occurrences}

We conducted additional analyses to determine which characteristics of an ECG recording influence the forecasting of impending AF using the given data.

\subsubsection{Heart Rate and RMSSD}\label{sssec:heart_rate_rmssd}

\begin{figure}[htbp]
\centering
\includegraphics[width=\textwidth]{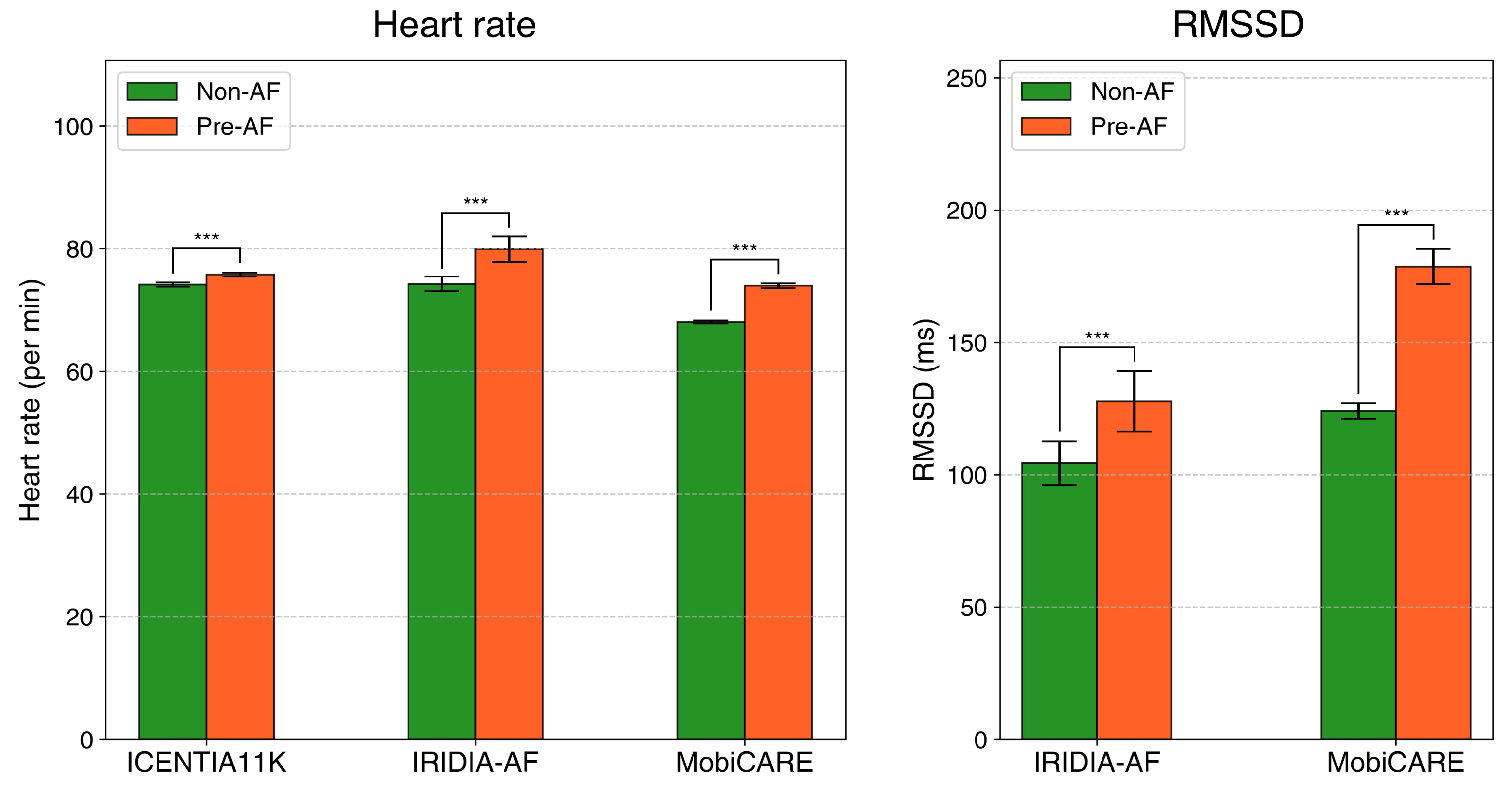}
\caption{
Comparison of heart rate and RMSSD between non-AF and pre-AF episodes across ICENTIA11K, IRIDIA-AF, and MobiCARE datasets.
Mean values with 95\% CI are shown.
Significance level: ***$p<0.01$.
}
\label{fig:hr_rmssd}
\end{figure}

Previous studies have shown that heart rate acceleration followed by an increase in RMSSD (Root Mean Square of Successive RR interval Differences) reflects a characteristic feature before AF onset, because sympathetic activation raises heart rate, and a subsequent surge in vagal activity elevates heart rate variability ~\citep{chen2014role, oh2024neuromodulation}.
To investigate whether these features differ between non-AF and pre-AF episodes, we calculated heart rate and RMSSD for patients in the personalization sets using their patient-specific test data.
Each feature was computed over 60-second segments using the Python library NeuroKit2 ~\citep{Makowski2021neurokit}. However, for the ICENTIA11K dataset, RMSSD values were unrealistically high (often exceeding 350 ms) in both episode types due to noise in the recordings. Therefore, RMSSD from ICENTIA11K was excluded from the analysis. The results are presented in Figure~\ref{fig:hr_rmssd}.

Compared to the non-AF dataset, the pre-AF dataset had significantly increased heart rates across the databases; in non-AF episodes, they are 74.2, 74.3, and 68.1 beats per minute for ICENTIA11K, IRIDIA-AF, and MobiCARE, respectively, while they increase to 75.8, 79.9, and 74.0 beats per minute in pre-AF episodes (all $p$-values $<0.01$). Similarly, the pre-AF dataset showed significantly increased RMSSD than the non-AF dataset; RMSSD values for non-AF episodes are 104.4 ms and 124.1 ms for IRIDIA-AF and MobiCARE, respectively, rising to 127.7 ms and 178.7 ms in pre-AF episodes (all $p$-values $<0.01$). Based on these results, we concluded that heart rate and RMSSD increased significantly in pre-AF episodes regardless of databases, with higher heart rate and RMSSD serving as potential indicators of imminent AF occurrences.

%%%%%%%%%%%%%%%%%%%%%%%%%%%%%%%%%%%%%%%%%%%%%%%%%%%%%%%%%%%%
%%%%%%%%%%%%%%%%%%%%%%%%%%%%%%%%%%%%%%%%%%%%%%%%%%%%%%%%%%%%

\subsubsection{Feature Attribution Analysis
}\label{ssec:feature_attribution_analysis}

To interpret what the AF forecasting model had learned from the data, we performed a feature attribution analysis to identify which characteristics of an ECG recording influenced its prediction. Feature attribution is a technique that measures the importance of each region of an input for a model’s decision, thereby providing insights into model behavior in ECG analysis~\citep{jahmunah2022explainable, singh2022interpretation, kwon2024classification}.
Regions with higher attribution values can be interpreted as having greater importance for the model’s prediction.
Among various feature attribution methods, we employed the absolute value of Guided Grad-CAM~\citep{selvaraju2017grad}, which has been shown to be effective in various ECG diagnostic tasks~\citep{suh2024visual}.
We conducted the analysis using personalized models trained for each patient in the ICENTIA11K dataset, with detailed implementation procedures provided in \ref{sec:implementation_of_feature_attribution_analysis}.

\begin{figure}[htbp]
\centering
\includegraphics[width=\textwidth]{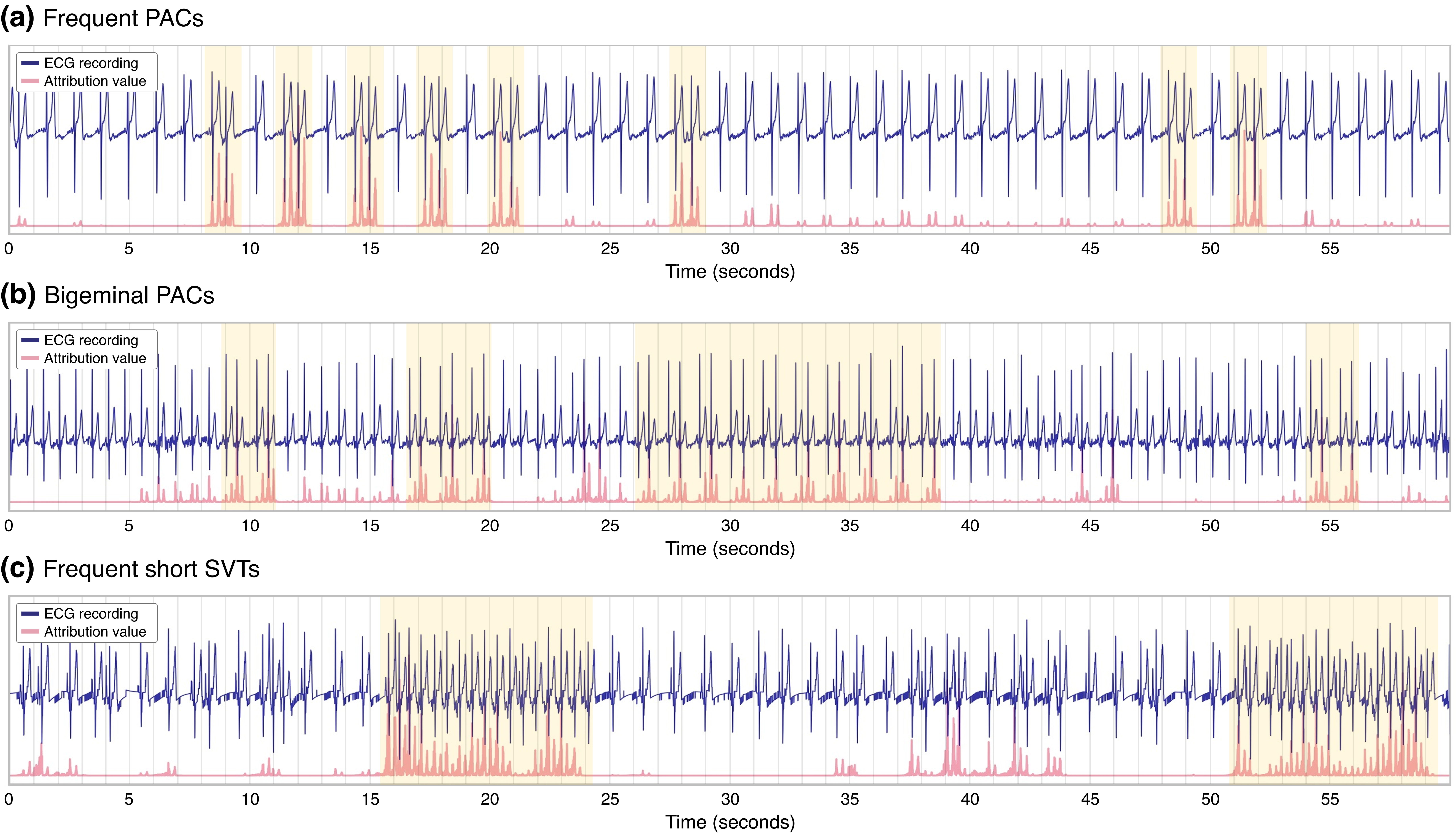}
\caption{
Visualization of feature attribution analysis for representative ECG characteristics preceding impending AF onset. The blue line represents the ECG recording, the red line indicates the attribution values for the model’s prediction of pre-AF episodes, and the yellow shaded regions mark characteristic features. (a) Frequent premature atrial complexes (PACs); (b) bigeminal PACs; (c) frequent, short supraventricular tachycardias (SVTs).
}
\label{fig:attribution}
\end{figure}

Figure~\ref{fig:attribution} illustrates representative ECG characteristics from samples predicted by the model as imminent AF, along with their corresponding attribution values.
Three arrhythmic patterns were identified: frequent premature atrial complexes (PACs), bigeminal PACs, and short supraventricular tachycardias (SVTs).
A higher burden of PACs and SVTs is known to be associated with an increased risk of developing AF.
Enhanced sympathetic and parasympathetic activity can promote atrial ectopic automaticity and reentry, thereby contributing to the development of tachyarrhythmias and leading to a greater burden of PACs and SVTs~\citep{chen2014role}.
The increasing burden of PACs and SVTs may also account for the observed increases in heart rate and RMSSD in pre-AF episodes before AF onset.
Notably, as shown in Figure~\ref{fig:attribution}, our model---trained solely on 60-second pre-AF and non-AF recordings---focused on PACs and SVTs when forecasting impending AF.
This alignment with established electrophysiological mechanisms supports the physiological plausibility and interpretability of our model.

%% file: tables/internal_validation_results.tex
\begin{table}[htbp]
  \centering
  \caption{%
    Internal validation results
    (ICENTIA11K, $n=30$ patients).
    Values were reported as mean\,[95\%\,CI].
    Significance levels: *$p<0.1$, **$p<0.05$.
  }
  \label{tab:internal_validation_results}

  \small      % keep the table compact; remove if full‐size text is preferred
  \begin{threeparttable}
    \begin{tabularx}{\textwidth}{@{}l *{2}{>{\centering\arraybackslash}X} c@{}}
      \toprule
      \textbf{Metric} &
      \textbf{Global model} &
      \textbf{Personalized model} &
      \textbf{$p$‐value} \\
      \midrule
      AUROC        & 0.614\,[0.551–0.678] & 0.711\,[0.652–0.770] & 0.0460\textsuperscript{**} \\
      AUPRC\tnote{a}
                   & 0.546\,[0.473–0.619] & 0.649\,[0.590–0.707] & 0.0212\textsuperscript{**} \\
      Sensitivity  & 0.725\,[0.662–0.788] & 0.728\,[0.680–0.777] & 0.9231 \\
      Specificity  & 0.643\,[0.554–0.731] & 0.730\,[0.670–0.790] & 0.1304 \\
      PPV          & 0.619\,[0.556–0.681] & 0.672\,[0.619–0.724] & 0.1619 \\
      NPV          & 0.766\,[0.718–0.813] & 0.812\,[0.766–0.859] & 0.1168 \\
      F1‐score     & 0.600\,[0.553–0.648] & 0.656\,[0.615–0.698] & 0.0653\textsuperscript{*} \\
      \bottomrule
    \end{tabularx}

    \begin{tablenotes}[flushleft]\footnotesize
      \item[a] A random‐guess classifier on this task yields an AUPRC of 0.377.
    \end{tablenotes}
  \end{threeparttable}
\end{table}

%% file: tables/external_validation_results_iridia-af.tex
\begin{table}[htbp]
  \centering
  \caption{%
    External validation results
    (IRIDIA-AF, $n=6$ patients).
    Values were reported as mean\,[95\%\,CI].
  }
  \label{tab:external_validation_results_iridia-af}

  \small
  \begin{threeparttable}
    \begin{tabularx}{\textwidth}{@{}l *{2}{>{\centering\arraybackslash}X} c@{}}
      \toprule
      \textbf{Metric} &
      \textbf{Global model} &
      \textbf{Personalized model} &
      \textbf{$p$‐value} \\
      \midrule
      AUROC        & 0.633\,[0.464–0.802] & 0.715\,[0.561–0.868] & 0.1954 \\
      AUPRC\tnote{a}
                   & 0.535\,[0.353–0.716] & 0.580\,[0.393–0.767] & 0.5335 \\
      Sensitivity  & 0.820\,[0.741–0.899] & 0.861\,[0.762–0.961] & 0.5820 \\
      Specificity  & 0.594\,[0.381–0.807] & 0.625\,[0.446–0.805] & 0.6521 \\
      PPV          & 0.594\,[0.446–0.743] & 0.601\,[0.432–0.770] & 0.9117 \\
      NPV          & 0.789\,[0.666–0.912] & 0.853\,[0.698–1.008] & 0.2106 \\
      F1‐score     & 0.638\,[0.553–0.722] & 0.677\,[0.533–0.821] & 0.5677 \\
      \bottomrule
    \end{tabularx}

    \begin{tablenotes}[flushleft]\footnotesize
      \item[a] A random‐guess classifier on this task yields an AUPRC of 0.333.
    \end{tablenotes}
  \end{threeparttable}
\end{table}

%% file: tables/external_validation_results_mobicare.tex
\begin{table}[htbp]
  \centering
  \caption{%
    External validation results
    (MobiCARE, $n=31$ patients).
    Values were reported as mean\,[95\%\,CI].
    Significance levels: *$p<0.1$, **$p<0.05$, ***$p<0.01$.}
  \label{tab:external_validation_results_mobicare}

  \small
  \begin{threeparttable}
    \begin{tabularx}{\textwidth}{@{}l *{2}{>{\centering\arraybackslash}X} c@{}}
      \toprule
      \textbf{Metric} &
      \textbf{Global model} &
      \textbf{Personalized model} &
      \textbf{$p$‐value} \\
      \midrule
      AUROC        & 0.585\,[0.524–0.645] & 0.686\,[0.629–0.742] & 0.0179\textsuperscript{**} \\
      AUPRC\tnote{a}
                   & 0.484\,[0.431–0.536] & 0.557\,[0.495–0.618] & 0.0517\textsuperscript{*}  \\
      Sensitivity  & 0.668\,[0.602–0.733] & 0.775\,[0.719–0.831] & 0.0051\textsuperscript{***}\\
      Specificity  & 0.625\,[0.543–0.707] & 0.635\,[0.556–0.714] & 0.8569                       \\
      PPV          & 0.553\,[0.506–0.599] & 0.576\,[0.522–0.630] & 0.4612                       \\
      NPV          & 0.743\,[0.682–0.804] & 0.853\,[0.827–0.880] & 0.0077\textsuperscript{***}\\
      F1‐score     & 0.538\,[0.497–0.580] & 0.628\,[0.584–0.671] & 0.0015\textsuperscript{***}\\
      \bottomrule
    \end{tabularx}

    \begin{tablenotes}[flushleft]\footnotesize
      \item[a] A random‐guess classifier on this task yields an AUPRC of 0.333.
    \end{tablenotes}
  \end{threeparttable}
\end{table}

%% file: TEX/50.discussion.tex
\section{Discussion}\label{sec:discussion}

In this study, we found that fine-tuning a deep learning model with a small amount of patient-specific wearable ECG data improves short-term forecasting of impending atrial fibrillation compared with a population-trained global model. Personalization improved performance on ICENTIA11K, yielded clear gains on MobiCARE despite its different lead configuration, and produced numerical improvements on IRIDIA-AF, which includes fewer patients. Our analyses further show that increasing the number of patient episodes in the fine-tuning set generally amplifies the benefit of personalization. We also observed that performance for the global model rises as the forecasting window approaches AF onset, whereas personalized models display different temporal dynamics in two of the three datasets, likely because they capture patient-specific patterns. In addition, pre-AF episodes exhibit higher heart rate and greater HRV measured by RMSSD, and the model emphasizes clinically plausible precursors such as premature atrial complexes (PACs) and supraventricular tachycardias (SVTs).

Our findings also demonstrate that the effectiveness of personalization is linked to patient-specific data density.
While fine-tuning yielded significant performance gains over a global model when a sufficient number of episodes were available for adaptation, the model showed only numerical improvements or became prone to overfitting when training samples were sparse.
This highlights the clinical necessity of a data-aware policy for model deployment.
In practice, we suggest an automated system that initially provides population-level forecasting via a global model and transitions to personalized fine-tuning only once a sufficient number of patient-specific episodes have been accumulated and annotated. Based on the trends in Figure~\ref{fig:effect_n_training_samples}, on the order of a dozen episodes may serve as a practical starting threshold, although the exact value should be calibrated to the deployment setting.
% In practice, we suggest an automated system where the system initially provides population-level forecasting via a global model, only transitioning to personalized fine-tuning once a robust statistical threshold, such as 12 or more patient-specific episodes, has been reached and annotated.
This ensures that personalization is only activated when the likelihood of performance enhancement outweighs the risk of generalization error.

While absolute discrimination leaves room for improvement, personalization yielded consistent and statistically significant gains across cohorts and device types. Population-based models assume homogeneous ECG patterns yet face substantial inter-patient variability---a known cause of inconsistent performance in unseen individuals---so these results indicate that patient-specific fine-tuning is important for building reliable, clinically useful ECG-based AI.

% While absolute discrimination leaves room for improvement, personalization yielded consistent and statistically significant gains across cohorts and device types, and this study offers valuable clinical insights for developing ECG-based AI systems. Our findings indicate that individualized ECG characteristics strongly influence model behavior, emphasizing the need for personalization in clinical applications. Population-based deep learning models often assume homogeneous ECG patterns across patients, yet inter-patient variability can be substantial, leading to inconsistent performance in unseen individuals. Our results suggest that such limitations can be mitigated through personalized fine-tuning. Therefore, incorporating personalization is essential for optimizing model accuracy and reducing variability across patients in clinically useful ECG-based AI models.

Unlike previous studies that focused on early warning of AF within 2 weeks~\citep{singh2022short, gadaleta2023prediction}, we aimed to develop a deep learning algorithm to forecast an impending AF within 5 minutes. Because of these different timescales, each study may emphasize different clinical utilities. For example, alerting individuals to an increased risk of AF within 2 weeks may encourage lifestyle changes to mitigate AF risk.
However, on a shorter timescale, as in our study, the algorithm may be more useful in critical-care settings where real-time monitoring and timely intervention are essential.
In the intensive care unit, patients' medical status can change dynamically, and AF with rapid ventricular rates can be critical and potentially lethal, necessitating prompt clinical decisions.
In such settings, short-term forecasting may be more useful than long-term forecasting~\citep{hinrichs2024short}, as it can help physicians promptly assess patient status and decide whether to administer anti-arrhythmic drugs.
Realizing this in practice will require further gains in forecasting precision to minimize false alarms and unnecessary medication use---an important direction for subsequent work.

Consistent with prior work~\citep{gavidia2024early, singh2022short, gadaleta2023prediction}, forecasting performance declined as the window moved farther from onset (Section~\ref{ssec:forecasting_future_af}). However, because those studies forecast over horizons of 30 minutes to two weeks, direct numerical comparison with our 5-minute results would be misleading, and the relevant ECG precursors likely differ across timescales.

% Existing studies using RR-interval features or ECG morphology have reported meaningful early warning performance, with accuracy declining as the time before AF onset increases~\citep{gavidia2024early, singh2022short, gadaleta2023prediction}. Similarly, our study observed a decline in performance as the prediction time point moved further away from AF onset. However, direct comparisons between our results and those of prior studies should be avoided. Our model targets AF forecasting within a 5-minute horizon, whereas previous studies evaluated predictions 30 minutes or up to 2 weeks in advance~\citep{gavidia2024early, singh2022short, gadaleta2023prediction}. As these studies operate on fundamentally different timescales, direct performance comparisons can be misleading. Moreover, the ECG features associated with upcoming AF attacks may differ across time scales.

While heart rate and RMSSD were analyzed to provide physiological context, an HRV-based classifier was not implemented as a baseline due to our 60-second window size and noise in the ICENTIA11K dataset, which resulted in unreliable HRV parameters except for heart rate. To calculate reliable frequency-domain parameters such as high, low, and very low frequencies, segments longer than the 60-second window are required. Future research using higher-fidelity recordings and different experimental settings could benefit from comparing raw ECG-based models with traditional HRV-based classifiers to further demonstrate the incremental value of deep learning in capturing morphological arrhythmic patterns.

Our results highlight significant inter-device heterogeneity. The ICENTIA11K, IRIDIA-AF, and MobiCARE datasets vary not only in sampling rates (200–256 Hz) but also in lead configurations and recording durations. This heterogeneity introduces a domain shift that challenges global models. Specifically, the global model struggled with the MobiCARE dataset, likely due to morphological discrepancies between Lead I (the training source) and Lead II (the target). While both leads are used for AF detection, Lead I reflects horizontal vectors, whereas Lead II captures the heart's primary electrical axis, typically yielding higher-amplitude P-waves and different QRS complexes. Therefore, the ECG features of Lead I that the global model utilized to distinguish pre-AF from non-AF episodes may not be identically represented in Lead II.

Furthermore, annotation granularity varies across datasets. As detailed in \ref{sec:af_annotation_lengths}, IRIDIA-AF focuses on long, contiguous AF spans (median $>1$ hour), while ICENTIA11K and MobiCARE capture much shorter paroxysmal events (median $\sim20$ seconds). These differences in ``ground truth'' labels mean that a global model may not be optimized for the specific requirements of how an AF detection or forecasting model should behave across diverse clinical contexts.

Within a continuous monitoring workflow, personalization acts as a recalibration step: fine-tuning on a small amount of patient-specific data adapts the model to both individual physiology and the device's hardware-specific signal morphology, bridging inter-patient and inter-device gaps. This suggests that for ECG-based AI to be portable, deployment strategies must shift from static global architectures toward adaptive frameworks that account for unique patient and device signatures.

%%% Limitations
There are several limitations to this work, primarily related to data availability.
First, the small size of the IRIDIA-AF dataset limits the interpretability and generalizability of the results.
The personalization set for IRIDIA-AF was restricted to only six out of the original 167 patients, as most recordings did not meet the criteria for personalization.
This small sample size introduces potential selection bias and provides insufficient coverage to generalize how the model adapts to the wide variety of individual ECG signal characteristics found in the broader population.
Furthermore, such a limited dataset size reduces statistical power and makes the results highly sensitive to individual episodes.
This limitation is further compounded by the absence of patient identifiers; without these, we cannot guarantee that the recordings represent unique individuals, which potentially narrows the diversity of our external validation set.
Second, we capped the number of episodes per patient at 30 to balance patients and classes, a constraint that may have limited the full potential of patient-specific training.
Allowing all available episodes might have further enhanced personalization.
Third, our analysis was limited to five-minute episodes with a 20-minute margin for non-AF episodes because ICENTIA11K is provided in 70-minute segments, preventing evaluation of longer windows or earlier forecasting horizons.
Fourth, restricting our analysis to patients with an AF burden below 60\% could impact the generalizability of the findings. We prioritized this subgroup because forecasting is more beneficial for infrequent events and minimizes unnecessary alarms. While this aligns with real-world data---where the individuals with high AF burdens are less common than those with low AF burdens in paroxysmal AF populations~\citep{kwon2022comparison}---it potentially restricts the model’s applicability to patients with high-frequency episodes.

%%%%%%%%%%%%% Future works
While our study establishes the feasibility of a personalized deep learning framework for short-term AF forecasting, several directions remain for enhancing its technical robustness.
Since our primary goal was to validate the personalization framework itself, we employed a model based on ResNet---a commonly used model in ECG processing.
Our findings demonstrate that personalized fine-tuning can be a robust strategy for mitigating inter-patient variability and adapting to hardware-related discrepancies.
Consequently, evaluating more advanced architectures, such as recent Transformer-based models~\citep{vaswani2017attention}, represents a promising direction to further optimize forecasting accuracy.
To further stabilize model performance in small-data regimes, future work should investigate advanced adaptation techniques including gradual unfreezing, weight regularization, and lower-rank adaptation (LoRA)~\citep{hu2022lora}.
Furthermore, future research could explore hybrid frameworks that integrate ECG signals with external environmental context---such as physical activity, movement, or stress levels. Incorporating such contextual data, as seen in the CACHET-CADB~\citep{kumar2022cachet} approach, could help reduce false positives caused by activity-induced artifacts.

The development of personalized AF forecasting models is currently constrained by the limited availability of publicly accessible, long-term ambulatory ECG databases that include AF onset annotations.
Due to this limitation, our study relied on the ICENTIA11K and IRIDIA-AF public datasets, supplemented by a private MobiCARE cohort.
Future work should aim to assemble larger, longer, and publicly accessible datasets with AF annotations.
Utilizing ECG databases recorded over extended durations would allow for a more rigorous assessment of how performance changes over time following personalization.
To further advance this framework, subsequent studies are warranted to evaluate its clinical utility and diagnostic efficacy using real-world data. Specifically, prospective, multi-center cohort studies utilizing long-term ECG data from various wearable devices are required.
Such designs would enable further validation of how personalization enhances performance across different hardware, alongside multi-device platforms~\citep{machorro2023cloud} and cost-effectiveness analysis.

% Such designs would enable further validation of how personalization enhances performance across different hardware.
% In parallel, a universal platform capable of analyzing ECG data from multiple manufacturers is necessary to comprehensively evaluate model efficacy. The development of a cloud-based platform integrating various wearable devices would significantly facilitate this research~\citep{machorro2023cloud}.
% Lastly, to establish clinical viability, the cost-effectiveness of AI-assisted ECG monitoring for early forecasting of AF must be investigated.

%% file: TEX/60.conclusion.tex
\section{Conclusion}\label{sec:conclusion}
We demonstrated that adapting a global model with a small amount of patient-specific wearable ECG data improves the short-term forecasting of impending atrial fibrillation. The model's performance generally increased as more patient episodes became available for adaptation. Furthermore, the personalized models in two of the three cohorts exhibited temporal patterns distinct from the global model, suggesting that they may capture patient-specific early cues. Physiological summaries and feature attribution analyses confirmed clinical expectations, identifying several key precursors: higher heart rate, greater RMSSD, and an increasing burden of premature atrial complexes (PACs) and short supraventricular tachycardias (SVTs).

Our personalization framework is well-suited for real-time, ambulatory ECG monitoring: by adapting a global model with patient-specific data, it improves forecasting accuracy and enables timely alerts for preventive intervention. These findings underscore the importance of personalization in ECG-based AI systems. Future prospective studies using larger datasets and longer horizons will be essential to confirm the framework's reliability and clinical utility at scale.

% Our personalization framework is well-suited for real-time, ambulatory ECG monitoring. By adapting a global model with patient-specific data, it enhances AF forecasting accuracy, enabling timely alerts for patients and clinicians to initiate preventive interventions.
% These findings highlight the importance of personalization in developing ECG-based AI systems. Improved forecasting accuracy through patient-specific adaptation can facilitate timely treatment and promote overall well-being within routine wearable workflows.
% The same approach can be extended to other sensor-rich domains that collect continuous, individual-level signals. Future prospective studies using larger ECG datasets and longer forecasting horizons will be essential to confirm the reliability and clinical utility of this framework at scale.

%% file: TEX/70.additional.tex
\section*{Author Contributions}\label{sec:author_contributions}
J.S., S.K, and W.R. conceived and designed the experiments.
J.S. performed the experiments.
J.S. and S.K. analyzed the experimental results.
J.S., S.K., J.K., and W.R. wrote the paper.
Y.K.K. and H.S.S. provided the experimental data.
E.-K.C. and W.R. coordinated and supervised the project.

\section*{Data Availability}\label{sec:data_availability}
ICENTIA11K and IRIDIA-AF are public datasets available at the following website: ICENTIA11K~({\small\url{https://physionet.org/content/icentia11k-continuous-ecg/1.0}}); IRIDIA-AF~({\small\url{https://zenodo.org/records/8405941}}).
MobiCARE is a clinical study dataset and is not publicly available to protect patient privacy.

\section*{Code Availability}\label{sec:code_availability}
The source code for this study is available under the MIT License at {\small\url{https://github.com/SNU-DRL/Personalized-AF-Forecasting}}. The repository is structured as follows:
\begin{itemize}
    \item \texttt{/dataset}: Scripts for preprocessing raw ECG recordings into the structured, episode-based format used in our experiments.
    \item \texttt{/scripts}: Command-line scripts to execute the training, personalization, and evaluation pipelines.
    \item \texttt{/src}: Core implementation details, including ResNet-based architectures, loss functions, and optimization logic.
\end{itemize}
Detailed guides for dataset acquisition and reproduction of the experimental procedures are also provided in the repository.

\section*{Ethics Statement}\label{sec:ethics_statement}
The data collection for the MobiCARE dataset was approved by the Seoul National University Hospital Institutional Review Board (IRB No: 2006-224-1138, date of approval: 26 June 2020), adhering to the Declaration of Helsinki (revised 2013). All participants provided informed consent for their data to be used for research purposes, and the data was fully de-identified prior to its use in this study. The other datasets used (ICENTIA11K, IRIDIA-AF) are public datasets available for public access online.

\section*{Acknowledgements}\label{sec:acknowledgments}
This work was supported by the Korea Medical Device Development Fund grant funded by the Korea government (the Ministry of Science and ICT, the Ministry of Trade, Industry and Energy, the Ministry of Health \& Welfare, the Ministry of Food and Drug Safety) (Project Number: 2710037350, RS-2020-KD000173).

\section*{Declaration of Competing Interest}\label{sec:competing_interests}
Y.K.K. and H.S.S. are affiliated with SEERS Technology Co., Ltd, the developer of the mobiCARE\textsuperscript{TM} device used for data collection in this study. E.-K.C. has received research grants or speaking fees from Abbott, Bayer, BMS/Pfizer, Biosense Webster, Chong Kun Dang, Daewoong Pharmaceutical Co., Daiichi-Sankyo, DeepQure, Dreamtech Co., Ltd., Jeil Pharmaceutical Co., Ltd., Medtronic, Samjinpharm, Samsung Electronics Co., Ltd., SEERS Technology Co., Ltd., Skylabs, and Yuhan Corporation, and is a stockholder in SEERS Technology Co., Ltd.
The other authors declare no competing interests.

%% file: TEX/A1.implementation_details.tex
\setcounter{table}{0}
\renewcommand\thetable{A.\arabic{table}}
\renewcommand*{\theHtable}{\thetable}
\setcounter{figure}{0}
\renewcommand\thefigure{A.\arabic{figure}}
\renewcommand*{\theHfigure}{\thefigure}

\section{Implementation Details}\label{sec:implementation_details}

\begin{figure}[htbp]
\centering
\includegraphics[width=0.5\textwidth]{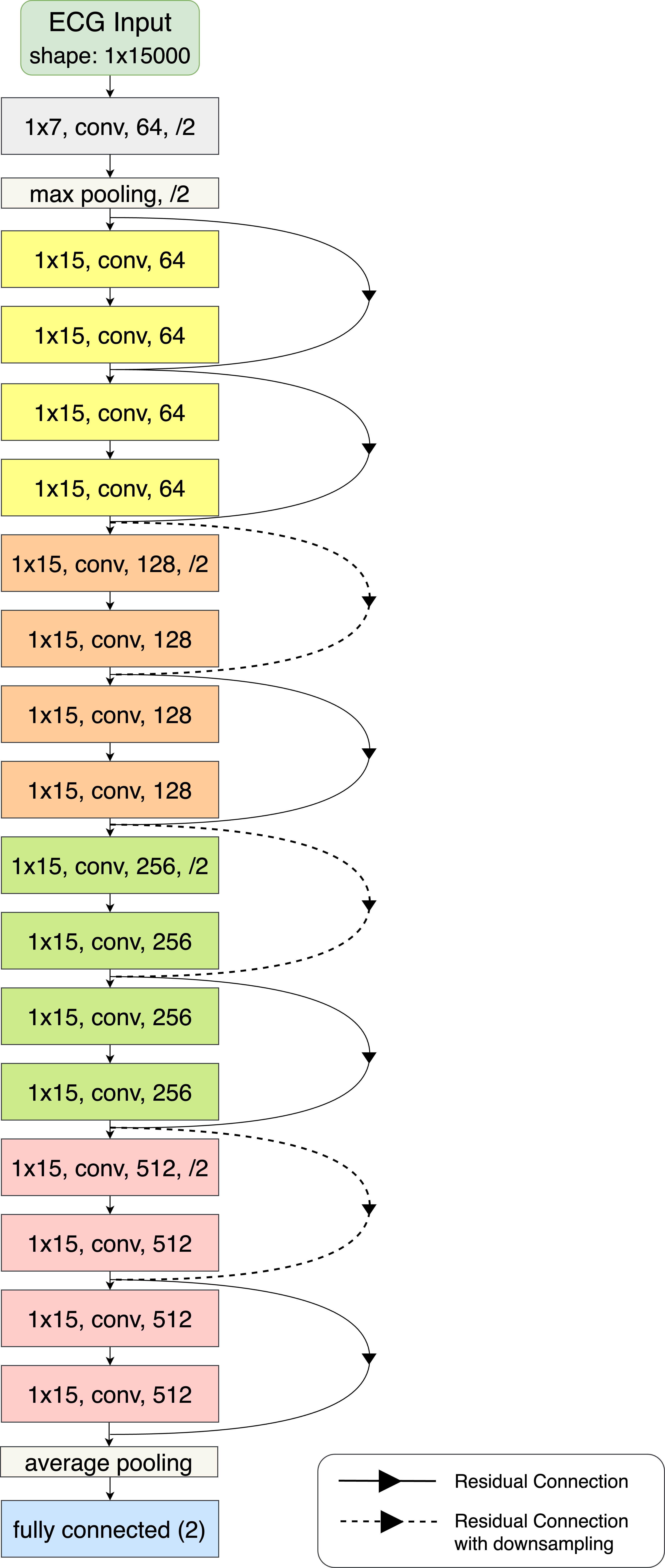}
\caption{
Architecture of the deep learning model used for AF forecasting.
}
\label{fig:model_architecture}
\end{figure}

We used the PyTorch~\citep{paszke2019pytorch} library to implement, train, and evaluate deep learning models.
The model, a modified 18-layer ResNet with 1x15 sized kernels designed for 1D ECG signal processing, is illustrated in Figure~\ref{fig:model_architecture}.
The general model was trained on the ICENTIA11K training and validation sets using cross-entropy loss with the Adam optimizer.
A cosine annealing scheduler adjusted the learning rate after each epoch, with a weight decay of $1 \times 10^{-4}$ and a batch size of 128.
To select the optimal configuration, we performed a grid search over learning rates $\{5\times10^{-2}, 1\times10^{-2}, 5\times10^{-3}, 1\times10^{-3}, 5\times10^{-4}, 1\times10^{-4}\}$ and training epochs $\{3, 5\}$.
The setting of a learning rate of $5e-3$ and 3 epochs yielded the best AUROC on the validation set and was adopted for subsequent experiments.

To train and evaluate the personalized models, we used the patient-specific training and test sets from ICENTIA11K, IRIDIA-AF, and MobiCARE.
The optimizer, learning rate scheduler, and weight decay were kept the same as in the general model, but the batch size was reduced to 32 to account for the smaller training sets.
We conducted a hyperparameter search over learning rates $\{1\times10^{-3}, 5\times10^{-4}, 1\times10^{-4}, 5\times10^{-5}\}$ and training epochs $\{3, 5, 10\}$. The selected configurations were $1\times10^{-3}$ with 5 epochs for ICENTIA11K, $1\times10^{-4}$ with 10 epochs for IRIDIA-AF, and $1\times10^{-3}$ with 5 epochs for MobiCARE.

For the experiments in Section~\ref{ssec:influence_of_training_episode_count}, we used the same hyperparameter settings as in the main experiments for ICENTIA11K and MobiCARE. The augmentation schemes described in Section~\ref{ssec:data_augmentation} were also applied. All reported metrics are the average of five runs with different random seeds.

%% file: TEX/A2.implementation_of_feature_attribution_analysis.tex
\setcounter{table}{0}
\renewcommand\thetable{B.\arabic{table}}
\renewcommand*{\theHtable}{\thetable}
\setcounter{figure}{0}
\renewcommand\thefigure{B.\arabic{figure}}
\renewcommand*{\theHfigure}{\thefigure}

\section{Implementation of Feature Attribution Analysis}\label{sec:implementation_of_feature_attribution_analysis}
We used the Captum~\citep{kokhlikyan2020captum} library, an open-source toolkit for deep learning interpretability, to implement feature attribution methods.
The analysis was conducted on personalized models trained for each patient in ICENTIA11K.
Predictions were made on the patient-specific test sets, and samples with a predicted probability of imminent AF of at least 70\% were selected.
Guided Grad-CAM was then applied using activations from the last convolutional layer of the model.
When using Guided Grad-CAM, we kept the parameter values at their default settings.
The absolute values of the resulting attribution maps were visualized alongside the corresponding ECG recordings.

%% file: TEX/A3.AF_annotation_lengths.tex
\setcounter{table}{0}
\renewcommand\thetable{C.\arabic{table}}
\renewcommand*{\theHtable}{\thetable}
\setcounter{figure}{0}
\renewcommand\thefigure{C.\arabic{figure}}
\renewcommand*{\theHfigure}{\thefigure}

\section{Distribution of AF Durations Across Databases}\label{sec:af_annotation_lengths}

Table~\ref{tab:af_lengths} describes the distribution of annotated AF durations across the three databases used in this study: ICENTIA11K, IRIDIA-AF, and MobiCARE.
IRIDIA-AF exhibits substantially longer AF spans, with a median ($Q_2$) duration of 4,924.7 seconds; this is more than 200 times longer than the medians observed in ICENTIA11K (21.3 seconds) and MobiCARE (19.7 seconds).
Overall, the annotations in ICENTIA11K and MobiCARE appear to capture brief, paroxysmal events or more granular episodes, whereas those in IRIDIA-AF encompass much broader temporal spans.
This tendency toward longer, contiguous annotations results in fewer total AF instances within IRIDIA-AF; across the 167 patients, the average is 2.32 AF spans per patient, where 127 individuals have only one or two recorded spans.
Consequently, these differences in label granularity result in significantly fewer pre-AF episodes being generated for IRIDIA-AF compared to other datasets, which explains the lower episode counts reported in Table~\ref{tab:num_episodes_personalization}.

\input{tables/af_lengths}

%% file: tables/af_lengths.tex
\begin{table}[htbp]
\centering
\caption{Statistical distribution of annotated AF episode durations (seconds). $P_{1}$ denotes the 1st percentile; $Q_{1}$, $Q_{2}$, and $Q_{3}$ represent the first, second (median), and third quartiles, respectively.}
\label{tab:af_lengths}
\begin{tabular}{@{}lrrrrrrr@{}}
\toprule
 \textbf{Dataset} & \textbf{Min} & \textbf{$P_{1}$} & \textbf{\textbf{$Q_{1}$}} & \textbf{\textbf{$Q_{2}$}} & \textbf{\textbf{$Q_{3}$}} & \textbf{Max} & \textbf{Mean} \\ \midrule
ICENTIA11K & 0.3  & 3.8  & 9.9   & 21.3   & 59.7    & 4\,194.3   & 99.4    \\
IRIDIA-AF  & 20.9 & 33.0 & 467.1 & 4\,924.7 & 14\,712.0 & 196\,241.2 & 14\,929.0 \\
MobiCARE   & 1.2  & 1.7  & 7.3   & 19.7   & 56.9    & 604\,800.0 & 1\,646.7  \\ \bottomrule
\end{tabular}%
\end{table}

%% file: TEX/A4.comparison_with_personalization_from_scratch.tex
\setcounter{table}{0}
\renewcommand\thetable{D.\arabic{table}}
\renewcommand*{\theHtable}{\thetable}
\setcounter{figure}{0}
\renewcommand\thefigure{D.\arabic{figure}}
\renewcommand*{\theHfigure}{\thefigure}

\section{Impact of Global Pre-training on Personalized Model Performance}\label{sec:comparison_with_personalization_from_scratch}

In our framework, the global model serves as a pre-trained foundation that captures generalizable population-level ECG patterns, while the personalized model leverages this transferred knowledge to adapt to an individual’s unique cardiac characteristics. To validate the necessity of this pre-training and fine-tuning paradigm, we compared the performance of our personalized model (fine-tuned from the global model) against a model trained from scratch using only individual patient-specific data. This analysis evaluates whether the features learned from a diverse population provide a superior starting point for adaptation compared to learning solely from the limited data typically available in a single patient's personalization set. The model, hyperparameter search process, and other training details are identical for both settings, as explained in \ref{sec:implementation_details}.

As shown in Table~\ref{tab:scratch_internal_validation_results}, global pre-training provides a substantial advantage for personalization in the ICENTIA11K internal validation dataset ($n=30$). The fine-tuned model achieved significantly higher AUROC (0.711 vs. 0.624, $p=0.0044$) and AUPRC (0.649 vs. 0.554, $p=0.0012$) compared to the model trained from scratch. While sensitivity remained comparable between the two approaches, the fine-tuned model demonstrated a significant improvement in specificity (0.730 vs. 0.629, $p=0.0113$), indicating that pre-training on a large dataset helps the model better distinguish between pre-AF episodes and non-AF episodes.

\input{tables/scratch_internal_validation_results}

For the external cohorts, the benefits of pre-training were more modest in MobiCARE or less evident in IRIDIA-AF. In the IRIDIA-AF dataset ($n=6$) (Table~\ref{tab:scratch_external_validation_results_iridia-af}), the fine-tuned model showed numerical improvements in AUROC (0.715 vs. 0.705), sensitivity (0.861 vs. 0.734), and F1-score (0.677 vs. 0.661), but other metrics exhibited numerical decreases, making it difficult to conclude the fine-tuned model's superiority in this specific, small cohort.
For the MobiCARE dataset ($n=31$) (Table~\ref{tab:scratch_external_validation_results_mobicare}), the fine-tuned model outperformed the scratch model in both AUROC (0.686 vs. 0.653) and AUPRC (0.557 vs. 0.517), although the significance of this effect was smaller than that seen in ICENTIA11K. We believe the reduced effectiveness of pre-training in these external cohorts is due to domain discrepancies between the pre-training source and target datasets. As detailed in \ref{sec:af_annotation_lengths}, IRIDIA-AF annotations reflect substantially longer AF durations and different patterns of granularity. Furthermore, MobiCARE utilizes a modified lead II configuration, whereas the global model was trained on lead I recordings from ICENTIA11K.

\input{tables/scratch_external_validation_results_iridia-af}

\input{tables/scratch_external_validation_results_mobicare}

We utilized only ICENTIA11K for global training due to its extensive size, so the impact of global pre-training was not as noticeable for IRIDIA-AF and MobiCARE. However, if diverse data from the two cohorts were available for the global model training phase, we anticipate the effect of pre-training would be more significant. In real-world clinical applications, pre-training with diverse, pre-collected data followed by fine-tuning with newly collected patient remains a robust strategy for enhancing the reliability of personalized AF forecasting.

%% file: tables/scratch_internal_validation_results.tex
\begin{table}[htbp]
  \centering
  \caption{%
    Internal validation results
    (ICENTIA11K, $n=30$ patients).
    Values were reported as mean\,[95\%\,CI].
    Significance levels: *$p<0.1$, **$p<0.05$, ***$p<0.01$.
  }
  \label{tab:scratch_internal_validation_results}

  \small      % keep the table compact; remove if full‐size text is preferred
  \begin{threeparttable}
    \begin{tabularx}{\textwidth}{@{}l *{2}{>{\centering\arraybackslash}X} c@{}}
      \toprule
      \textbf{Metric} &
      \textbf{Personalized Model (Scratch)} &
      \textbf{Personalized Model (Fine-tuned)} &
      \textbf{$p$‐value} \\
      \midrule
      AUROC        & 0.624\,[0.551-0.698] & 0.711\,[0.652–0.770] & 0.0044\textsuperscript{***} \\
      AUPRC\tnote{a}
                   & 0.554\,[0.492-0.617] & 0.649\,[0.590–0.707] & 0.0012\textsuperscript{***} \\
      Sensitivity  & 0.733\,[0.668-0.798] & 0.728\,[0.680–0.777] & 0.8615 \\
      Specificity  & 0.629\,[0.537-0.721] & 0.730\,[0.670–0.790] & 0.0113\textsuperscript{**} \\
      PPV          & 0.623\,[0.568-0.679] & 0.672\,[0.619–0.724] & 0.0679\textsuperscript{*} \\
      NPV          & 0.776\,[0.708-0.844] & 0.812\,[0.766–0.859] & 0.1299 \\
      F1‐score     & 0.631\,[0.585-0.677] & 0.656\,[0.615–0.698] & 0.1329 \\
      \bottomrule
    \end{tabularx}

    \begin{tablenotes}[flushleft]\footnotesize
      \item[a] A random‐guess classifier on this task yields an AUPRC of 0.377.
    \end{tablenotes}
  \end{threeparttable}
\end{table}

%% file: tables/scratch_external_validation_results_iridia-af.tex
\begin{table}[p]
  \centering
  \caption{%
    External validation results
    (IRIDIA-AF, $n=6$ patients).
    Values were reported as mean\,[95\%\,CI].
  }
  \label{tab:scratch_external_validation_results_iridia-af}

  \small
  \begin{threeparttable}
    \begin{tabularx}{\textwidth}{@{}l *{2}{>{\centering\arraybackslash}X} c@{}}
      \toprule
      \textbf{Metric} &
      \textbf{Personalized Model (Scratch)} &
      \textbf{Personalized Model (Fine-tuned)} &
      \textbf{$p$‐value} \\
      \midrule
      AUROC        & 0.705\,[0.574-0.837] & 0.715\,[0.561–0.868] & 0.7718 \\
      AUPRC\tnote{a}
                   & 0.591\,[0.409-0.773] & 0.580\,[0.393–0.767] & 0.7725 \\
      Sensitivity  & 0.734\,[0.528-0.940] & 0.861\,[0.762–0.961] & 0.1228 \\
      Specificity  & 0.760\,[0.588-0.931] & 0.625\,[0.446–0.805] & 0.1362 \\
      PPV          & 0.697\,[0.514-0.881] & 0.601\,[0.432–0.770] & 0.1822 \\
      NPV          & 0.868\,[0.781-0.956] & 0.853\,[0.698–1.008] & 0.7746 \\
      F1‐score     & 0.661\,[0.506-0.815] & 0.677\,[0.533–0.821] & 0.2283 \\
      \bottomrule
    \end{tabularx}

    \begin{tablenotes}[flushleft]\footnotesize
      \item[a] A random‐guess classifier on this task yields an AUPRC of 0.333.
    \end{tablenotes}
  \end{threeparttable}
\end{table}

%% file: tables/scratch_external_validation_results_mobicare.tex
\begin{table}[htbp]
  \centering
  \caption{%
    External validation results
    (MobiCARE, $n=31$ patients).
    Values were reported as mean\,[95\%\,CI].
    Significance levels: *$p<0.1$.}
  \label{tab:scratch_external_validation_results_mobicare}

  \small
  \begin{threeparttable}
    \begin{tabularx}{\textwidth}{@{}l *{2}{>{\centering\arraybackslash}X} c@{}}
      \toprule
      \textbf{Metric} &
      \textbf{Personalized Model (Scratch)} &
      \textbf{Personalized Model (Fine-tuned)} &
      \textbf{$p$‐value} \\
      \midrule
      AUROC        & 0.653\,[0.588-0.718] & 0.686\,[0.629–0.742] & 0.2429 \\
      AUPRC\tnote{a}
                   & 0.517\,[0.446-0.588] & 0.557\,[0.495–0.618] & 0.1562  \\
      Sensitivity  & 0.788\,[0.723-0.853] & 0.775\,[0.719–0.831] & 0.5830\\
      Specificity  & 0.575\,[0.481-0.670] & 0.635\,[0.556–0.714] & 0.1389                       \\
      PPV          & 0.535\,[0.482-0.588] & 0.576\,[0.522–0.630] & 0.0650\textsuperscript{*}  \\
      NPV          & 0.796\,[0.720-0.872] & 0.853\,[0.827–0.880] & 0.1526 \\
      F1‐score     & 0.603\,[0.555-0.651] & 0.628\,[0.584–0.671] & 0.1138 \\
      \bottomrule
    \end{tabularx}

    \begin{tablenotes}[flushleft]\footnotesize
      \item[a] A random‐guess classifier on this task yields an AUPRC of 0.333.
    \end{tablenotes}
  \end{threeparttable}
\end{table}

%% file: TEX/A5.additional_tables.tex
\setcounter{table}{0}
\renewcommand\thetable{E.\arabic{table}}
\renewcommand*{\theHtable}{\thetable}
\setcounter{figure}{0}
\renewcommand\thefigure{E.\arabic{figure}}
\renewcommand*{\theHfigure}{\thefigure}

\section{Additional Tables}\label{sec:additional_tables}

\input{tables/num_episodes_training_validation}
\input{tables/num_episodes_personalization}

%% file: tables/num_episodes_training_validation.tex
\begin{table}[htbp]
  \centering
  \caption{Number of 5-minute episodes and 60-second segments in the ICENTIA11K training and
           validation sets. Each episode was divided into 25 overlapping 60-second
           segments (10-second stride), which serve as the model inputs.}
  \label{tab:num_episodes_training_validation}

  \begin{threeparttable}
    \small                       % keeps the table compact—remove if not desired
    % l = left-aligned text; X = stretchable column (ragged-right here)
    \begin{tabularx}{\textwidth}{@{}ll*{2}{>{\raggedleft\arraybackslash}X}@{}}
      \toprule
      \multicolumn{2}{l}{} & \textbf{Training} & \textbf{Validation} \\ 
      \midrule
      \multicolumn{2}{l}{Patients} & 173 & 43 \\ 
      \midrule
      \multirow{2}{*}{5-min episodes} & Total        & 1\,800 & 499 \\ 
                                & Per patient  & 10.4 ± 8.7 & 11.6 ± 10.8 \\ 
      \midrule
      \multirow{2}{*}{60-sec segments} & Total       & 45\,000 & 12\,475 \\ 
                                     & Per patient & 260.1 ± 217.3 & 290.1 ± 271.2 \\ 
      \bottomrule
    \end{tabularx}

  \end{threeparttable}
\end{table}

%% file: tables/num_episodes_personalization.tex
\begin{sidewaystable}[p] % rotates 90° and floats on its own page
  \centering
  \caption{Number of 5-minute episodes and 60-second segments in the personalization sets.
            Each episode was divided into 25 overlapping 60-second
           segments (10-second stride), which served as model inputs. Identical patient pools were used for the patient-specific training and
          test sets within each dataset.}
  \label{tab:num_episodes_personalization}

  \footnotesize
  \begin{threeparttable}
  \begin{tabularx}{\linewidth}{@{}l l
               *{2}{>{\raggedleft\arraybackslash}X}
               *{2}{>{\raggedleft\arraybackslash}X}
               *{2}{>{\raggedleft\arraybackslash}X}@{}}
    \toprule
    \multicolumn{2}{l}{} &
      \multicolumn{2}{c}{\textbf{ICENTIA11K}} &
      \multicolumn{2}{c}{\textbf{IRIDIA-AF}} &
      \multicolumn{2}{c}{\textbf{MobiCARE}} \\
    \cmidrule(lr){3-4}\cmidrule(lr){5-6}\cmidrule(lr){7-8}
    \multicolumn{2}{l}{} &
      \textbf{Patient-specific} & \textbf{Patient-specific} &
      \textbf{Patient-specific} & \textbf{Patient-specific} &
      \textbf{Patient-specific} & \textbf{Patient-specific} \\
    \multicolumn{2}{l}{} &
      \textbf{training set} & \textbf{test set} &
      \textbf{training set} & \textbf{test set} &
      \textbf{training set} & \textbf{test set} \\
    \midrule
    \multicolumn{2}{l}{Patients}                  & 30 & 30 & 6 & 6 & 31 & 31 \\ 
    \midrule
    \multirow{2}{*}{5-min episodes} & Total             & 561 & 632 & 42  & 57  & 576 & 744 \\
                              & Per patient       & 18.7 ± 8.9 & 21.1 ± 7.9
                                                    & 7.0 ± 1.4 & 9.5 ± 5.9
                                                    & 18.6 ± 8.4 & 24.0 ± 10.2 \\
    \midrule
    \multirow{2}{*}{60-sec segments} & Total         & 14\,025 & 15\,800 & 1\,050 & 1\,425 & 14\,400 & 18\,600 \\
                                  & Per patient   & 467.5 ± 221.6 & 526.7 ± 196.4
                                                    & 175.0 ± 35.4 & 237.5 ± 146.3
                                                    & 464.5 ± 209.0 & 600.0 ± 255.6 \\
    \bottomrule
  \end{tabularx}

  \end{threeparttable}
\end{sidewaystable}